\documentclass[12pt]{article}
\usepackage{amsmath}
\usepackage{graphicx,psfrag,epsf}
\usepackage{enumerate}
\usepackage{booktabs}
\usepackage{url} 
\usepackage[style=authoryear]{biblatex}
\newcommand{\blind}{1}

\usepackage[english]{babel}
\usepackage{csquotes}

\usepackage[colorinlistoftodos]{todonotes}
\usepackage[colorlinks=true, allcolors=black]{hyperref}
\usepackage{amsfonts,amssymb,mathrsfs}
\usepackage{mathtools}
\usepackage{algorithm}
\usepackage[noend]{algpseudocode}
\usepackage[font=footnotesize,labelfont=bf]{caption}

\begin{document}

\def\spacingset#1{\renewcommand{\baselinestretch}%
{#1}\small\normalsize} \spacingset{1}


\if1\blind
{
  \title{\bf Adaptive Mantel Test for Association Testing in Imaging Genetics Data}
  \author{
    Dustin Pluta\hspace{.2cm}\\
    Department of Statistics, University of California, Irvine\footnote{Corresponding author: Dustin Pluta, Department of Statistics, University of California, Irvine CA 92627 (e-mail: dpluta@uci.edu).}\\
    Hernando Ombao\hspace{.2cm}\\ 
    King Abdullah University of Science and Technology\\
    Chuansheng Chen \hspace{.2cm}\\
    Department of Psychological Science, University of California, Irvine\\
    Gui Xue\hspace{.2cm}\\ 
    Center for Brain and Learning Sciences, Beijing Normal University\\
    Robert Moyzis\hspace{.2cm}\\
    Department of Biological Chemistry, University of California, Irvine\\
    Zhaoxia Yu\hspace{.2cm}\\
    Department of Statistics, University of California, Irvine}
    \date{}
  \maketitle
} \fi

\if0\blind
{
  \bigskip
  \bigskip
  \bigskip
  \begin{center}
    {\LARGE\bf Adaptive Mantel Test for Association Testing in Imaging Genetics Data}
\end{center}
  \medskip
} \fi

\bigskip
\newpage

\begin{abstract}
An important goal of imaging genetics studies is the identification of neuroimaging phenotypes that are significantly associated with an individual's genotype. To improve on existing methods for association-testing, we here propose the adaptive Mantel test, a flexible method for testing the association of two sets of high-dimensional features simultaneously across multiple similarity metrics, without the need to directly adjust the significance threshold for multiple testing. In this paper, we develop the theoretical properties of the adaptive Mantel test through a unified framework of linear model score tests, which links the fixed effects and variance components models via the ridge regression score test.  Simulation results show that the adaptive Mantel test has higher power than existing association testing methods in many settings.  Using the proposed methods to analyze data from a novel imaging genetics study of visual working memory in healthy adults, we identified interesting associations of brain connectivity (measured by EEG coherence) with selected genetic features.  Supplementary materials, including an R-package for the adaptive Mantel test, are available online.
\end{abstract}

\noindent%
\vfill
\newpage

\spacingset{1.45} 
\begin{center}\section{INTRODUCTION}\label{introduction}\end{center}
Imaging genetics studies typically collect neuroimaging and genetic data with hundreds of thousands of features per modality for each subject, while sample sizes are usually on the order of $1000$. Given how little is known about the relationship of neurological phenotypes with genetics, it is of interest to determine whether particular sets of features are significantly associated in aggregate. Developing association tests that are sufficiently powered in high-dimensional settings remains an important statistical challenge for the analysis of imaging genetics data, which will be addressed in this paper.  

We here propose the adaptive Mantel test (AdaMant), a flexible method for testing the association of two sets of high-dimensional features.  AdaMant is designed to simultaneously test over multiple similarity measures such that the correct type I error rate under the null hypothesis is maintained without the need to directly adjust the significance threshold for multiple testing, and is able to achieve higher power than alternative methods by data-adaptively selecting the similarity measure that minimizes the test $P$-value. In particular, we shall illustrate AdaMant by testing the association between genetic features and brain connectivity measured by EEG coherence in a novel imaging genetics data set collected by our co-authors.

Several parametric and non-parametric approaches to multi-modal association testing have been developed and revisited in the scientific and statistics literature, but these approaches generally suffer from low power when the feature sets are high-dimensional relative to the sample size. The random effects model has been widely used to handle correlated data such as repeated measurements in longitudinal studies \parencite{laird1982random}. Because it can naturally account for correlation across observations, the variance components form of the random effects model has been used for many years in genetic studies to estimate the heritability of phenotypes from a sample of related or unrelated subjects \parencite{kwee2008powerful, tzeng2009gene, schaid2010genomic, pan2011relationship, goeman2004global, goeman2006testing, liu2007semiparametric, yang2011gcta}. In particular, the score test of the variance components model has been used for testing the association of two high-dimensional sets of features due to the ease of calculating the test statistic \parencite{ge2016multidimensional}.

Mantel's test \parencite{mantel1967} is one of the earliest formulations of a (ostensibly non-parametric) distance-based test for association of features from two observational modalities.  In the Mantel test, one computes a similarity or distance between each pair of subjects in each modality, and tests for significance of the correlation of similarities.  Closely related tests have been developed from the RV coefficient \parencite{robert1976unifying} and the distance covariance \parencite{szekely2007measuring}, and the sum of powered score tests \parencite{xu2017adaptive}. A detailed discussion of the connections of MT, the RV coefficient, and distance covariance is given in \parencite{omelka2013comparison}.
Equivalent methods have frequently been used for genetic association analysis, e.g.  \cite{schork2012statistical, wessel2006generalized,salem2010curve}.


Recent related work by \citeauthor{liu2018geometric} (\citeyear{liu2018geometric}) gives a detailed examination of association tests constructed from combinations of principal components, motivated by limitations of the Wald test and canonical correlation analysis for multivariate association testing.  \citeauthor{liu2018geometric} consider different methods for weighting the combination of PCs to achieve higher power than competing methods in some settings.  While the weighted PC tests developed are a useful extension of existing methods, the existing literature still lacks a general framework that relates existing tests and allows for the construction and interpretation of new association testing methods.

From the perspective of the Mantel test, the present work develops such a framework encompassing the RV coefficient test, weighted combination of PCs, canonical correlation analysis, and many other association tests.  The primary contribution of this paper is the development of the adaptive Mantel test (AdaMant), a generalization of the Mantel test designed to leverage the advantages of ridge regression for high-dimensional testing.  To address the difficulty in selecting ridge tuning parameters, AdaMant allows one to simultaneously test across a pre-selected set of parameters, with only a small loss in power compared to knowing the optimal parameter \textit{\`a priori}. From a convenient form of the AdaMant test statistic related to the principal components of the covariates, we show that AdaMant has higher power than the RV and dCov tests for certain choices of data generating mechanisms.  In developing the theoretical properties of AdaMant, we additionally show that the score tests of the fixed effects, ridge regression, and variance components model form a one-parameter class of association tests.  

The structure of  this article is as follows.  In Section \ref{score_tests} we develop a framework of linear model score tests and show that the linear regression score test is equivalent to the Mantel test for particular choices of similarity measures. Section \ref{correlations} presents theoretical results on correlations of features in the context of the kernel Mantel test, and relates these correlations to the proportion of variance explained and signal-to-noise ratio.  Section \ref{adamant} describes the implementation and use of the adaptive Mantel test.  Section \ref{simulations} evaluates the performance of the test for univariate response data, and for simulated, realistic imaging genetics data. Section \ref{application} presents results showing significant associations between brain connectivity (as measured by EEG coherence) and genetic features from a working memory study of 350 college students.  Section \ref{discussion} concludes with a discussion of generalizations and directions for future work.  Additional details and results of this article are available in the online supplement.  An R-package providing a simple implementation of AdaMant is available on the author's website \if1\blind{\parencite{pluta2018}}\fi.

\section{LINEAR MODEL SCORE TESTS AND THE MANTEL TEST}
\label{score_tests}

The testing procedure we will develop in this paper is based on the linear model score test, which is ideal for high-dimensional inference since it does not require estimation of the parameters to be tested, greatly reducing the computation required compared to alternatives such as the Wald test and likelihood ratio test.  This also makes it convenient to calculate the null distribution for the score test statistic via a permutation procedure, which forms the basis of the adaptive Mantel test algorithm proposed in Section \ref{adamant}.  

In this section, we derive convenient expressions for the score tests for the fixed effects, variance components, and ridge regression models as weighted norms of the principal correlation vector of the feature sets.  Using these forms of the score test statistics, we give a novel unification of these three linear model score tests as a one-parameter class of association tests, parameterized by the ridge penalty term $\lambda \in [0, \infty]$, where $\lambda = 0$ corresponds to the fixed effects score test, and $\lambda = \infty$ corresponds to the variance components score test.  We then establish the equivalence of the kernel Mantel test and the linear model score tests for particular choices of metrics, which further shows that the linear model score tests form an important subclass of the more general class of similarity-based association tests derived from weighted Euclidean inner products.  We conclude this section by extending these association tests to settings with multivariate response data, and discusses the implications of using the ridge kernel in this setting.

\subsection{A Unified Framework of Linear Model Score Tests}
\label{linear_model_score_tests}

For data $(X_i, Y_i) \in \mathbb{R}^p \times \mathbb{R}^q, i = 1, \dots, n$, we consider a stochastic model with high-dimensional parameter vector of interest \(\beta\).  We denote the likelihood as \(\mathcal{L}(\beta)\), score vector \(\mathcal {U}(\beta) = \frac{\partial \log \mathcal L}{\partial \beta}\), and Fisher information \(\mathcal I(\beta) = -\mathbb{E} \frac{\partial}{\partial \beta} \mathcal U(\beta)\). The score test statistic for testing $H_0: \beta = \beta_0$ is $S=\mathcal U(\beta)^T(\mathcal I(\beta))^{-1}\mathcal U(\beta)|_{\beta=\beta_0}$ (replacing any nuisance parameters by consistent estimators). 

We now derive expressions for the score test of the association of $X$ and $Y$ in the fixed effects, ridge regression, and variance components models, and show that these form a single class of linear association tests indexed by the ridge penalty $\lambda \in [0, \infty]$. Throughout this paper, we assume $X$ has been column-centered to zero mean and $Y$ has been standardized to zero mean and unit variance.  The score test statistics can be compactly written by considering the singular value decomposition (SVD) of $X$.  Assuming $\text{rank}(X) = r \leq \min(p, n)$, the SVD is $X = UDV^T$, where $U$ is $n \times r$ with $U^TU = I_r$, $D$ is an $r \times r$ diagonal matrix of singular values, where $\eta_j, j = 1, \cdots, r$ are the squared singular values, and $V$ is $p \times r$, with $V^TV = I_r$.  We refer to the columns of $U$ as the \textit{principal directions} of $X$.  It is convenient to also define $\eta_j = d_j^2$, which is the variance of $X$ along the $j$th principal direction, and  the \textit{principal correlation vector} $Z = U^TY$, which is an $r \times 1$ vector with $j$th component $Z_j = R(Y, U_j)$, the Pearson correlation of $Y$ with $X$ along the $j$th principal direction.  When two test statistics $T_1$, $T_2$ differ by values that are fixed under permutation of the subject labels (which do not affect the permutation $P$-value), we say that $T_1$ is \textit{testing equivalent} to $T_2$, written $T_1 \asymp T_2$.

\subsubsection{Fixed Effects Linear Regression}

The classical fixed effects linear regression model can be written $Y = X\beta+\epsilon$, where $\varepsilon\sim N(0,\sigma_{\varepsilon}^2 I_n).$ This model is broadly applicable for simple association testing, but requires $n > p$, and so is not feasible for high-dimensional settings. Assuming that $X$ is full column rank ($\text{rank}(X) = p$), the score test statistic for the fixed effects model is equivalent to 

\begin{equation}T_F \asymp Y^TX(X^TX)^{-1}X^TY = \text{tr}(YY^TX(X^TX)^{-1}X^T)) = \sum_{j = 1}^p Z_j^2,
\label{eqn:score_F}
\end{equation}

where $Z_j$ is the $j$th component of $Z = U^TY$, i.e. the correlation of $Y$ with the $j$th principal component of $X$.\\

\subsubsection{Variance Components Model}

The variance components model can be stated as $Y = Xb + \varepsilon$, where \(b \sim N(\mathbf{0}, \sigma_b^2I_p)\), \(\varepsilon \sim N(\mathbf{0}, \sigma_e^2I_n)\). The null hypothesis of no linear association between $X$ and $Y$ corresponds to $\sigma_b^2=0$. Similar to \citeauthor{liu2007semiparametric} (\citeyear{liu2007semiparametric}), we first calculate the score and use the term that involves both $X$ and $Y$ as our score test statistic, which yields the equivalent form

\begin{equation}
T_V \asymp Y^TXX^TY = \text{tr}(YY^TXX^T) = \sum_{j = 1}^r\eta_j Z_j^2.
\label{eqn:score_V}
\end{equation}

In the context of genetic heritability analysis, the variance components model is usually reparameterized as

\begin{equation}
Y = g + \varepsilon,
\label{eqn:vc}
\end{equation} 

\noindent where $g \sim N(0, \sigma^2 XX^T / p)$.  The random effects vector $g$ can be interpreted as the aggregate effect of the features in $X$, with $\sigma^2 = p\sigma^2_b$.  In the sequel we will write this model as $Y \sim N(0, \sigma^2 XX^T / p + \sigma^2_{\varepsilon}I_n)$.

\noindent \textbf{Remark} $T_F$ and $T_V$ are both sums of the squared principal correlations, with $T_V$ weighting each term by the variance ($\eta_j$) of $X$ along that principal component.  Thus, $T_V$ gives greater emphasis to those directions with higher variance, whereas $T_F$ gives equal weight to each direction. To better understand the relationship of these tests, we next consider the ridge regression score test statistic, which provides a natural link between $T_F$ and $T_V$.\\

\subsubsection{Ridge Regression}

The ridge regression model is a $L^2$-penalized form of the fixed effects model, for which the estimator of \(\beta\) is defined as the minimizer of the penalized likelihood, or equivalently $\hat\beta_{\lambda} = {\arg \min}_{b} \left\{||Y-Xb||^2+\lambda||b||^2\right\}$.  Alternatively, ridge regression can be formulated as an augmented data model, $Y_\lambda = X_\lambda \beta + \epsilon$ where $Y_{\lambda}=(Y^T | ~\mathbf{0}_{1 \times p})^T, X_\lambda = (X^T | ~\sqrt{\lambda} I_p)^T$.  From the likelihood for this augmented model, the ridge regression score statistic is equivalent to

\begin{align}
T_\lambda \asymp \text{tr}(X_{\lambda}Y_{\lambda}Y_{\lambda}^TX_{\lambda}(X_{\lambda}^TX_{\lambda})^{-1}X_{\lambda}^T) = \sum_{j = 1}^r \frac{\lambda\eta_j}{\eta_j + \lambda} Z_j^2.
\label{eqn:score_lambda}
\end{align}

\subsubsection{Unifying the Linear Model Score Tests}\label{univariate_ridge}

Using the SVD of $X$, the null distributions for the three models can be derived by observing that $Z = U^TY \overset{H_0} \sim N_r(0, I_r),$ from which the distributions of the score test statistics follow from standard results on quadratic forms of normal random variables. Proposition 1 establishes the ridge regression score test as a link between the fixed effects and variance components score tests. In each case, the test statistic follows a mixture of chi-squared distributions (with one degree of freedom), which can be approximated by the Satterthwaite method (\cite{satterthwaite1946approximate}) or a Monte Carlo procedure.  Alternatively, the $P$-value can be calculated by permuting subject labels to compute the reference distribution.\\

\noindent \textbf{Proposition 1.} 

Let $X$ be a column-centered $n \times p$ matrix with $\text{rank}(r)$, SVD $X = UDV^T$, and squared singular values $\eta_j, j = 1, \cdots, r$.  Let $Y \sim N(0, \sigma^2 \Sigma_A + \sigma^2_{\varepsilon} I_n)$ be an $n \times 1$ response vector, where $\Sigma_A$ is an $n \times n$ semi-positive definite covariance matrix that depends on $X$.  Define $Z = U^TY$, the $r \times 1$ vector of principal correlations, and define $T_F, T_V$, and $T_{\lambda}$ by Equations \ref{eqn:score_F}, \ref{eqn:score_V}, and \ref{eqn:score_lambda}.

\begin{enumerate}
\item \textbf{Connections between linear model score tests.} The score tests for the fixed effects, variance components, and ridge regression models form a single class of tests parameterized by the ridge tuning parameter $\lambda \in [0, \infty]$. $T_{\lambda}$ is testing equivalent to the variance components score test as $\lambda \to \infty$, $\lim_{\lambda \to \infty} T_{\lambda} \asymp T_V.$ Assuming $\text{rank}(X) = p$, $T_{\lambda = 0}$ is testing equivalent to the fixed effects score statistic, $T_0 \asymp T_F$.  

\item \textbf{Score test reference distributions.}
Let $\chi_{1, j}^2, j = 1, \dots, r$ be iid chi-squared random variables with one degree of freedom.  Let $\chi_p^2$ be a chi-squared random variable with $p$ degrees of freedom.
\begin{enumerate}
\item $T_{\lambda} \asymp \sum_{j=1}^{r}\frac{\lambda\eta_j}{\lambda+\eta_j} Z_j^2  ~\overset{H_0} \sim ~ \sum_{j=1}^{r}\frac{\lambda\eta_j}{\lambda+\eta_j} \chi^2_{1, j}.$
\item $T_F \asymp \sum_{j = 1}^p Z_j^2 ~\overset{H_0} \sim ~ \chi_p^2.$  
\item $T_V \asymp \sum_{j = 1}^r \eta_j Z_j^2 ~\overset{H_0} \sim ~ \sum_{j = 1}^r \eta_j \chi_{1, j}^2.$
\end{enumerate}
\end{enumerate}

\noindent \textbf{Outline of Proof} (1-1) can be shown by directly evaluating the limit.  Since $Y$ has been standardized, (1-2) follows directly from $Z = U^TY \overset{H_0}{\sim} N(0, I_n)$ and $Z_j^2 \sim \chi_{1, j}^2$ for $\chi_{1, j}^2$ iid $j = 1, \cdots, r$. Complete proofs given in \ref{prop1}.\\

\noindent \textbf{Remark.} To better understand the differences between these score tests, we present a geometric interpretation of the test statistics. Recognizing $Z = U^TY$ as the vector of correlations of $Y$ along each principal direction of $X$, we can interpret the test of $T_F \asymp \|Z\|^2$ as testing the Euclidean norm of the correlation vector.  In contrast, the statistic $T_V \asymp \sum_{j = 1}^r \eta_j Z_j^2$, thus the variance components score test statistic is equivalent to testing the weighted norm of $Z$, where the $j$th component (corresponding to the $j$th principal direction of $X$) is weighted by $\eta_j$ (the variance of $X$ along the $j$th principal direction).  This has the effect of emphasizing the influence of directions in $\textbf{X}$ for which $X$ has large variance and reducing the influence of directions with small variance.  The \underline{ridge regression score test} is a compromise between the fixed effects and variance components tests, with small $\lambda$ yielding a test close to the fixed effects (and identical at $\lambda = 0$), and large $\lambda$ yielding a test close to the variance components score test, converging to identical tests as $\lambda \to \infty$.  In other words, the ridge test weights $Z_j$ proportional to $\eta_j$ as in the variance components test, but flattens each weight by a factor of $\frac{1}{\lambda + \eta_j}$.  

\subsection{Mantel Test}\label{mantel}

We next review the Mantel test \parencite{mantel1967} and show that it includes the score test considered above as a special case. The classical Mantel test, using the distance matrices for two sets of features, has been shown to exhibit extremely low power in even modestly high-dimensional settings, and so how largely fallen out of use.  However, the framework of the Mantel test is extremely general, and encompasses many well-known association tests, as we will show.

Let $X$ and $Y$ be as above.  For $\mathcal{K}^{\textbf{X}}(\cdot, \cdot)$ a symmetric positive semi-definite kernel (or similarity) function on $\textbf{X} \times \textbf{X}$, let $K$ be the corresponding Gram matrix with $K_{ij} = \mathcal{K}^{\textbf{X}}(X_i, X_j)$, and let kernel $\mathcal{K}^{\textbf{Y}}$ defined on $\textbf{Y} \times \textbf{Y}$ have corresponding Gram matrix $H$. The Mantel test statistic for these kernels is defined as

\[T(X, Y) = \text{tr}\left(HK\right) = \sum_{i = 1}^n \sum^n_{j = 1} H_{ij}K_{ij}.\]

Assuming observations are exchangeable, the reference distribution under the null hypothesis of
no association between $\mathcal{K}^{\textbf{X}}$ measures and $\mathcal{K}^{\textbf{Y}}$ measures, can be obtained from the observed features $X$ and $Y$ by permuting the observation labels for one set of features and calculating the empirical reference distribution for $T$. Equivalently, one can hold one matrix fixed, say \(H\), and simultaneously permute the rows and columns of \(K\).  The expected value of this test statistic under the null hypothesis is $\mathbb{E}[T(X, Y)] = \sigma^2_{\varepsilon}\text{tr}(K)$, which depends on the choice of $K$.

\begin{figure}
\centering
\includegraphics[scale = 0.5]{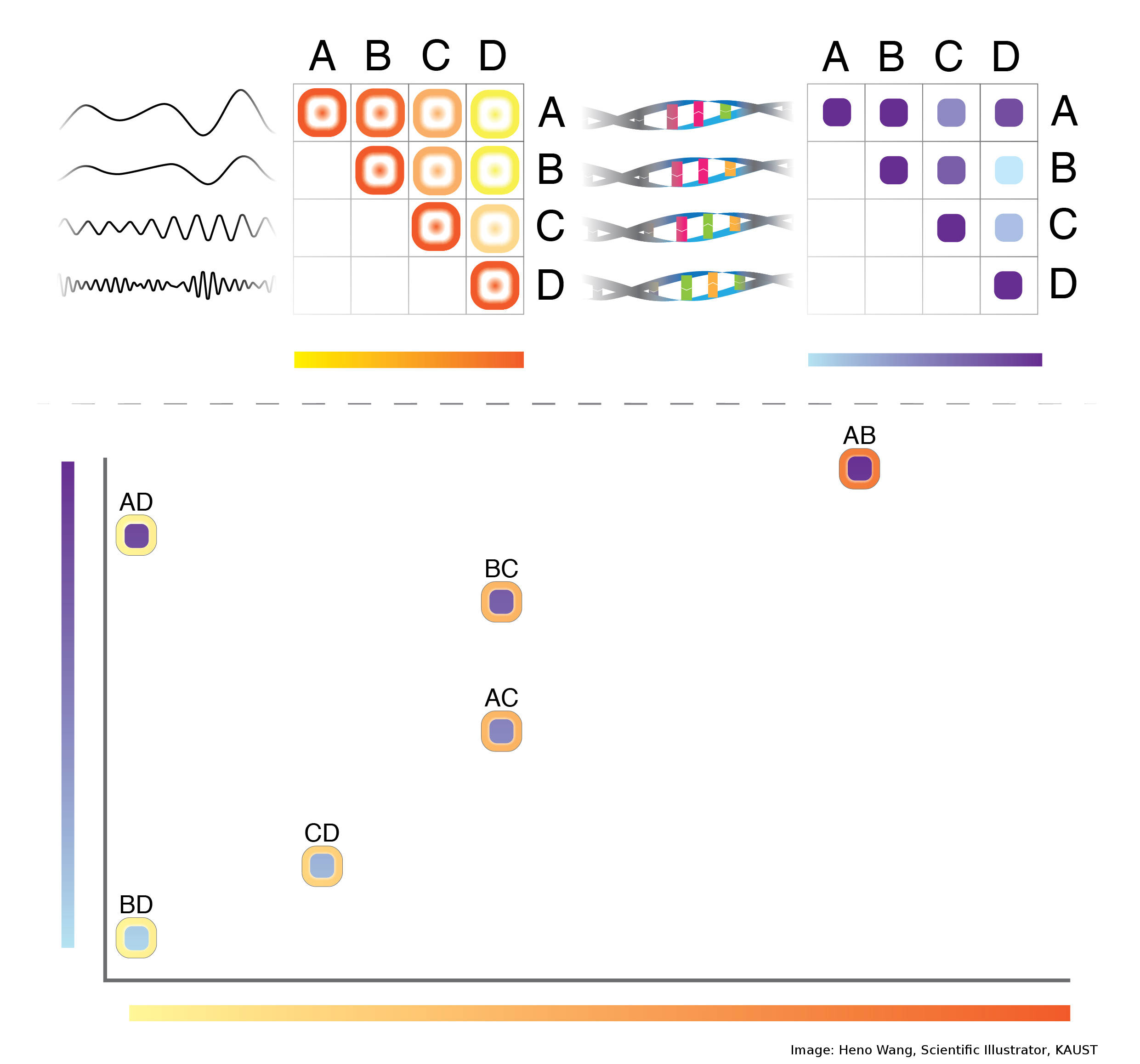}
\label{fig:mantel_diagram}
\caption{Conceptual illustration of applying the Mantel test to EEG and genetic data from subjects A-D.  (Top left) Gram matrix of filtered EEG data at a selected frequency band. Red indicates a high degree of similarity between EEG data from the indicated pair of subjects.  (Top right) Genetic relationship matrix. Purple indicates a high degree of genetic similarity between the indicated pair of subjects. \textit{(Bottom)} Similarity-by-similarity plot of the vectorized upper triangle of the above Gram matrices.  Points in the lower left region of the plot correspond to pairs of subjects that are dissimilar in both modalities.  Points in the upper right region correspond to pairs of subjects that are similar in both modalities.}
\end{figure}

We will focus on tests with similarity measures that can be written as $\mathcal{K}_{\mathcal{W}}(X_i, X_j) = \langle X_i, X_j\rangle_{\mathcal W} = X_i^T \mathcal{W} X_j$ for a positive semi-definite (p.s.d.) weight matrix $\mathcal W$, referred to as weighted Euclidean inner products.   We propose three choices for $\mathcal{W}$ and the resulting Gram matrices, which we denote $K_F, K_V,$ and $K_{\lambda}$, that produce Mantel tests equivalent to the score tests for the fixed effects, variance components, and ridge regression models respectively.  
Choosing $\mathcal W = I_p$ (the $p \times p$ identity matrix) gives the standard Euclidean inner product, with Gram matrix $K_V = XX^T$.  Another natural choice for weight matrix is $\mathcal{W} = (X^TX)^{-1}$ (assuming the inverse exists), which is the projection matrix into $\mathcal{C}(X)$, the column space of $X$.  The Gram matrix is $K_F = X(X^TX)^{-1}X^T.$  $K_F$ is also recognizable as the ``hat matrix'' from the fixed effects model, and is related to the Mahalanobis distance.  When $X^TX$ is not full rank, such as when $n > p$, we can replace $(X^TX)^{-1}$ with a generalized inverse $(X^TX)^{-}$, since the similarity matrix is invariant to the choice of inverse. Alternatively, as in ridge regression \parencite{hoerl1975ridge}, we can pre-condition the weight matrix by adding a positive constant $\lambda$ to the diagonal, i.e. \(\mathcal{W} = \lambda (X^TX + \lambda I_p)^{-1}\), which gives Gram matrix $K_{\lambda} = \lambda X(X^TX + \lambda I_p)^{-1}X^T.$  Letting $H = YY^T$, we define the Mantel test statistics $T_V, T_F, T_{\lambda}$ from these Gram matrices. Note that we use the same notation for the linear model score tests and corresponding Mantel tests since these tests are equivalent under a permutation testing procedure.\\

\noindent \textbf{Definition. Weighted Mantel Tests.}
\begin{align}
&\textbf{Euclidean:}~~ T_V = \text{tr}(HK_V) = Y^TXX^TY,\label{eqn:T_V}\\
&\textbf{Mahalanobis:}~~ T_F = \text{tr}(HK_F) = Y^TX(X^TX)^{-1}X^TY,\label{eqn:T_F}\\
&\textbf{Ridge Weighted:}~~ T_{\lambda} = \text{tr}(HK_{\lambda}) = \lambda Y^TX(X^TX + \lambda I_p)^{-1}X^TY
\end{align}

Since multiplying the test statistic by a term that does not depend on both $X$ and $Y$ does not affect the permutation $P$-value, it follows from the expressions for the score tests in Proposition 1 that the score tests for the fixed effects, variance components, and ridge regression models are equivalent to the Mantel test with the Mahalanobis, Euclidean, and ridge inner products respectively.  Thus the linear model score test statistics have the following equivalent interpretations: (i) the weighted norm of the correlation vector $X^TY$; (ii) the weighted sum of principal correlations; (iii) the correlation of similarities of subjects in \textbf{X} (measured by $\mathcal{K}_{\mathcal{W}}$) with the similarities of subjects in \textbf{Y} (measured by the Euclidean inner product).

\subsection{Mantel Test for Multivariate Outcomes}\label{multivariate}

Given the connection between fixed effects and variance components models, we now extend the preceding results to the relationship between two sets of multivariate measurements. This extension is 
essential for many applications, such as heritability analysis of neuroimaging-derived phenotypes. Previous results in multivariate modeling have shown that ridge regression with tuning parameters for both response and covariates can be effective for the simultaneous prediction of multiple outcome variables (\cite{brown1980adaptive,breiman1997predicting}), thus it may be useful to consider ridge kernel similarities for both feature sets in the Mantel test.

Suppose now that each subject $i$ is measured on a $q$-variate phenotype $Y_i$ so that $Y$ is an $n \times q$ matrix, and assume that $Y$ is column-centered, and let $X$ be a column-centered $n \times p$ covariate matrix as before.  For two tuning parameters $\lambda_X, \lambda_Y > 0$ and Gram matrices $H_{\lambda_Y}=Y(Y^TY+\lambda_Y I_q)^{-1}Y^T \mbox{ , } K_{\lambda_X}=X(X^TX+\lambda_X I_p)^{-1}X^T$, the ridge kernel Mantel test statistic is $T_{\lambda_X,\lambda_Y} = \text{tr}(H_{\lambda_Y}K_{\lambda_X})$.

To understand the underlying implications of the ridge kernel similarity measure, we consider a data-augmentation definition of ridge regression. Data augmentation has been widely used to ease the computation in several problems, such as the expectation and maximization algorithms, and Gibbs or MCMC sampling \parencite{duncan1972linear, dempster1977maximum, dempster1981estimation, tanner1987calculation, van2001art, gelman2014bayesian}. In particular, \citeauthor{hodges1998some} (\citeyear{hodges1998some}) described how data augmentation can be used as a device to write hierarchical models in the form of linear models. 

We demonstrate in \ref{multivariate_ridge} that $T_{\lambda_X, \lambda_Y}$ corresponds to a similar augmented data model as in the univariate response case.  Examining the log-likelihood, it is apparent that $\lambda_X$ shrinks $B$ toward zero, $\lambda_Y$ encourages smaller sums of the reciprocals of the eigenvalues of $\Sigma_{\varepsilon}$. Shrinking $B$ by the ridge penalization induced by $\lambda_X$ is equivalent to replacing $X(X^T+\lambda_X I_p)^{-1}X^T$ with $XX^T$ when $\lambda_X$ is large. When $\lambda_Y$ is large, we expect that $\Sigma_{\varepsilon}$ is approximately proportional to an identity matrix, that is $Y(Y^TY+\lambda_Y)^{-1}Y^T$ is approximately proportional to $YY^T$. Thus, large values of tuning parameters favors the use of Euclidean distance in that modality, which ignores the dependence structure of the features, leading to $\lim_{\lambda_X \to \infty, \lambda_Y \to \infty} T_{\lambda_X, \lambda_Y} = \text{tr}(XX^TYY^T)$. 


  \begin{center}
\section{CORRELATIONS AND VARIANCE EXPLAINED}
\label{correlations}
\end{center}
Correlation metrics and the proportion of variance explained are commonly used for quantifying the strength of association of $X$ and $Y$.  The product-moment (or Pearson) correlation is a symmetric measure of cosine similarity of the two random vectors, and is stated without reference to a particular model.  In contrast, the proportion of variance in $Y$ explained by $X$ is necessarily measured with respect to some model stated with $Y$ as response and $X$ as explanatory variables, and is thus not a symmetric measure.  Correlation is simple to calculate, but can only measure the linear relationship of $X$ and $Y$, and may be misleading if $X$ and $Y$ have a significant nonlinear relationship.  Measuring variance explained instead requires the selection and fitting of a model, allowing one to specify the functional relationship of $X$ and $Y$ to be measured.  We here develop relationships between these measures and the Mantel test statistics defined in the previous section.  

\subsection{Correlations}

The sample (Pearson) correlation of the similarity measures of all subjects can be naturally defined as

\begin{eqnarray}
R(H,K)=\frac{\text{tr}(HK)}{\sqrt{\text{tr}(H^2)\text{tr}(K^2)}}.
\end{eqnarray}

\noindent Since the trace is an inner product on the space of square matrices, \(R(H,K)\in [-1, 1]\) by the Cauchy-Schwarz Inequality, and for \(H\) and \(K\) SPD, this is further restricted to \(0 \leq R(H,K) \leq 1\). When permutations are
used to assess the strength of association, \(MT\) is equivalent
to testing the significance of \(R(H,K)\), since
the denominator of \(R(H,K)\) is fixed when
simultaneously permuting the rows and columns of either \(H\) or \(K\).

The quantities $R(K, H)$ and $\text{tr}(K, H)$ are related to known statistics for many choices of $K$ and $H$.  For example, $R(K_F, H_F)$ is proportional to Hooper's trace correlation $R_T^2$ \parencite{hooper1959simultaneous}, which is defined as  $R_T^2 = \frac{1}{q}\text{tr}\left((Y^TY)^{-1}Y^TX(X^TX)^{-1}X^TY)\right) = \sqrt{\frac{p}{q}}R(K_F, H_F)$.  This is also proportional to the trace of the canonical correlation matrix, and thus the test $T_{F,F} = \text{tr}(K_FH_F)$ can be interpreted as testing the sum of the squared canonical correlations.

The statistic $T_{F,V} = \text{tr}(X(X^TX)^{-1}X^TYY^T)$ for multivariate $X$ and $Y$ can be interpreted similar to the univariate case. Defining the principal correlations $Z_{jk} = U_j^TY_k$, which is the correlation of the $k$th feature of $Y$ along the $j$th principal direction of $X$.  Writing $T_{F, V} = \text{tr}(K_FH_V) = \sum_{j = 1}^r \sum_{k = 1}^s Z_{jk}^2$, this statistic can be understood as testing the Frobenius norm of the matrix of principal correlations.  Proposition 2 shows that the sample correlation of similarities for the three choices of $K$ above can be conveniently related through the singular value decomposition (SVD) of $X$ and the principal correlation vector.\\

\noindent \textbf{Proposition 2.} Let $X$ be an $n \times p$ column-standardized matrix of covariates with $\text{rank}(X) = r$ and singular value decomposition $X = U_{n\times r}D_{r\times r}V_{p\times r}^T$, with squared singular values $\eta_i, i = 1, \cdots, r$. Let $Y$ be an $n \times 1$ standardized vector of scalar responses, and let \(H=YY^T\) and $Z = U^TY$.
\begin{enumerate}
	\item 
	Fixed Effects
	\begin{equation}
		\label{eqn:fixed}
    	R(H, K_F) = \frac{\sum_{j=1}^r Z_j^2}{n\sqrt{p} } = \frac{1}{\sqrt{p}}R^2(X, Y) \le \frac{1}{\sqrt{p}}.
	\end{equation}
	\item Random Effects
	\begin{equation}
			\label{eqn:random}
		R(H, K_V) = \frac{\sum_{j=1}^r\eta_j Z_j^2}{n\sqrt{\sum_{j=1}^r\eta_j^2}}
        \end{equation}
    \item Ridge Regression
   	\begin{equation}
   			\label{eqn:ridge}
		R(H, K_{\lambda}) = \frac{\sum_{j=1}^{r} \frac{\eta_j}{\lambda + \eta_j}Z_j^2}{n\sqrt{\sum_{j=1}^{r} \left(\frac{\eta_j}{\eta_j+\lambda}\right)^2}}.
   	\end{equation}
   \item Limits of Multivariate Ridge Correlations.
   \begin{equation}
    \label{eqn:mv_corr_limit}
   		\lim _{\lambda_X\rightarrow \lambda_X^*, \lambda_Y\rightarrow \lambda_Y^*} R(H_{\lambda_Y},K_{\lambda_X})=R(H_{\lambda_Y^*}, K_{\lambda_X^*}),
   \end{equation}
        \noindent where $H_0 = H_F$, $K_0 = K_F$, $H_{\infty} = H_V$, $K_{\infty} = K_V$.   
\end{enumerate}

\noindent \textbf{Outline of Proof} Equations (\ref{eqn:fixed}) - (\ref{eqn:ridge}) follow from the definition of matrix correlation and the linear model score test expressions in Proposition 1.  The result in (\ref{eqn:mv_corr_limit}) follows from a direct computation of the limits.  Full details are given in \ref{prop2}.

\subsection{Proportion of Variance Explained}

The proportion of variance explained in both fixed-effect and random-effect models can be expressed as functions of Mantel statistics. In the genetics literature, $h^2$ is known as the additive, or \textit{narrow-sense}, genetic heritability \parencite{yang2011gcta, liu2007semiparametric}, defined as the proportion of phenotypic variance explained by additive SNP effects. Moment based estimators for $h^2$ have been proposed, such as the Haseman-Elston estimator , which regress pairwise distances in phenotypes to pairwise genetic correlation of related subjects \parencite{haseman1972investigation, elston2000haseman}. This has recently been adapted to unrelated subjects \parencite{yang2010common, golan2014measuring}, and extended to multivariate phenotypes as well \parencite{ge2016multidimensional}.  

To define heritability of multivariate phenotypes, we consider the variance components model with multivariate response $Y \sim N(0, \Sigma_A \otimes G + \Sigma_E \otimes I_n)$, where $\Sigma_A$ is the genetic covariance of the features of $Y$, $G = XX^T / p$, and $\Sigma_E$ is the covariance of the subject errors.  The proportion of variance explained by this model is defined as  

\begin{equation}
h^2=\frac{\text{tr}(\Sigma_A)}{\text{tr}(\Sigma_A) + \text{tr}(\Sigma_E)}.
\label{eqn:h2}
\end{equation}

In the simplified case of uncorrelated phenotypes with homogeneous genetic effect $\sigma^2$ and uncorrelated subjects with constant error variance $\sigma^2_{\varepsilon}$, the covariance matrices are $\Sigma_A = \sigma^2 I_q$ and $\Sigma_E = \sigma^2_{\varepsilon}$.  In this case, the heritability reduces to $h^2 = \sigma^2 / (\sigma^2 + \sigma^2_{\varepsilon})$.  The univariate variance components model (Equation \ref{eqn:vc}) is then recovered when $q = 1$.   The following proposition connects the correlation of Gram matrices with the proportion of variance explained in the multivariate variance components model.\\

\noindent \textbf{Proposition 3} Assume $X$ is an $n \times p$ column-standardized, full row-rank matrix (which requires $p \geq n$), and
assume $Y \sim N(0, \sigma^2 I_q \otimes XX^T / p + \sigma^2_{\varepsilon}I_q \otimes I_n)$ is a $q$-variate column-standardized random vector. Define Gram matrices $G = XX^T / p$ and $H = YY^T$ (for which $\text{tr}(G) = n$ and $\text{tr}(H) = nq$ under the above assumptions). 
\begin{enumerate}
	\item The method of moments estimator $\hat h^2$ is
	
	$$\hat h^2 = \frac{\text{tr}(HG / q) - n}{\text{tr}(G^2) - n}.$$

	\item The expectation of the correlation of Gram matrices is
		$$\mathbb{E} R(H, G) = h^2\frac{(\text{tr}(G^2) - n)}{\sqrt{\text{tr}(H^2 / q)\text{tr}(G^2)}} + \frac{n}{\sqrt{\text{tr}(H^2 / q)\text{tr}(G^2)}}.$$
\end{enumerate}
\vspace{1cm}

\noindent \textbf{Corollary} In the setting of Proposition 3, with $Y$ a univariate vector ($q = 1$) and for fixed $n$, $p$, and $\sigma^2$, with $p > n$, the expected correlation $\mathbb{E}R(H, G)$ is a function of $\text{tr}(G^2)$, with maximum and minimum dependent on the noise-to-signal ratio $(1 - h^2) / h^2$. 
	\begin{enumerate}
	    \item For $1 \leq (1 - h^2) / h^2 \leq n$, the expected correlation is bounded by
	    
	    $$2\sqrt{\frac{h^2(1 - h^2)}{n}} \leq \mathbb{E}R(H, G) \leq h^2\frac{n - 1}{n} + \frac{1}{n}.$$
	    
	    with upper and lower bounds attained by $\text{tr}(G^2) = n^2$ and $\text{tr}(G^2) = \frac{n(1 - h^2)}{h^2}$ respectively.
	    \item For $(1 - h^2) / h^2 \leq 1$, the expected correlation is bounded by
	    
	    $$\frac{1}{\sqrt{n}} \leq \mathbb{E}R(H, K) \leq h^2\frac{n - 1}{n} + \frac{1}{n},$$
	    
	    with upper and lower bounds attained at $\text{tr}(G^2) = n^2$ and $\text{tr}(G^2) = n$ respectively.
	    \item For $(1 - h^2) / h^2 \geq n$, the expected correlation is bounded by
	    
	    $$h^2\frac{n - 1}{n} + \frac{1}{n} \leq \mathbb{E}R(H, G) \leq \frac{1}{\sqrt{n}},$$
	    
	    with upper and lower bounds attained at $\text{tr}(G^2) = n$ and $\text{tr}(G^2) = n^2$ respectively.
	\end{enumerate}

\vspace{1cm}
\noindent \textbf{Outline of Proofs} (3-1) -- (3-2) are derived from applications of standard results on quadratic forms of normal random variables.  The corollary is determined by optimizing $\mathbb{E} R(H, G)$ with respect to $\text{tr}(G^2)$. Complete proofs given in \ref{prop3}.

\begin{center}\section{ADAPTIVE MANTEL TEST}\label{adamant}\end{center}
Effectively using kernel methods requires an appropriate selection of the kernel function and tuning parameters for the particular setting.  Selection methods have been extensively considered in the context of prediction problems, with 
cross-validation as the \textit{de facto} standard.  Cross-validation is a straight-forward and practical selection method for prediction, but may be difficult to implement for hypothesis testing, since the type I error rate needs to be controlled.  Furthermore, the tuning parameter selected by minimizing the CV MSE may not necessarily yield the highest powered test. In this section, we develop the adaptive Mantel test (AdaMant), a novel extension of the classical MT to utilize the ridge kernel in association testing methods.  By using the maximum $P$-value from a set of $T_{\lambda}$ statistics, AdaMant is able to simultaneously test across a set of tuning parameters and kernels without the need to directly apply adjustments for multiple tests. In comparison to the RV-test, which relies on a single metric, AdaMant tests across a set of metrics, and can therefore achieve higher power in certain settings, as shown by the simulations in Section \ref{simulations}.

\subsection{Algorithm for the Adaptive Mantel Test}

The ``adaptive'' procedure used here is similar to the adaptive sum of powered score test algorithm described in \citeauthor{xu2017adaptive} (\citeyear{xu2017adaptive}). The procedure receives as input a list of pairs of metrics/kernels $\{(\mathcal{K}^{\textbf{X}}_{m}, \mathcal{K}^{\textbf{Y}}_{m}) | m = 1 , \cdots M\}$ from which the matrices $K_m = \mathcal{K}^{\textbf{X}}_{m}(X)$ and $H_m = \mathcal{K}^{\textbf{Y}}_{m}(Y)$ are computed for each metric pair, $m = 1, \cdots, M$.  These metrics may be from a single family with varying tuning parameters, such as ridge kernels with different penalization terms, or may include kernels from different families.

For each $m = 1, \cdots, M$, $P_m$ is calculated as the $P$-value of the Mantel test with metrics $\mathcal{K}_m^{\textbf{X}}$ and $\mathcal{K}_m^{\textbf{Y}}$ for $X$ and $Y$ respectively.  The AdaMant test statistic is defined as the minimum of these values,

$$P^{(0)} := \min_{m \in \{1, \cdots, M\}} P_m.$$

A permutation procedure is practical for calculating the reference distribution for $P^{(0)}$.  For each $m$, and  $b = 1, \cdots, B$, $H_m^{(b)}$ is generated by permuting rows and columns of $H$ simultaneously, and the corresponding test statistic $P^{(b)}$ is calculated.  The AdaMant $P$-value is then calculated as

$$P_{AdaMant} = \frac{1}{B + 1} \sum_{b = 0}^B I\left(P^{(0)} \leq P^{(b)}\right).$$

\noindent General pseudocode for AdaMant is given in Algorithm 1.\\  


\begin{algorithm}
\caption{Adaptive Mantel Test}\label{alg:adamant}
\begin{algorithmic}[1]
\Procedure{AdaMant}{$X, Y, \{\mathcal{K}_m^{\textbf{X}}, \mathcal{K}_m^{\textbf{Y}}\}_{m = 1}^{M}$, B}
\State $K_m \gets \mathcal{K}^{\textbf{X}}_{m}(X)~~\forall~m = 1, \dots, M$
\State $H_m \gets \mathcal{K}^{\textbf{Y}}_{m}(Y)~~\forall~m = 1, \dots, M$
\State Calculate $T_{m}^{(0)} \gets T_{m} := \text{tr}(K_mH_m)$
\State Generate $B$ permutations of $H_m$, labeled $H_m^{(b)}~~\forall~m = 1, \dots, M; b = 1, \dots, B$.
\State $T_{m}^{(b)} \gets \text{tr}(K_{m}H_m^{(b)})~~\forall~m = 1, \dots, M; b = 1, \dots, B$
\State $P_{m}^{(b)} \gets \frac{1}{B + 1}\sum_{b = 0}^B I\left(Z_{m}^{(b)} \leq Z_{m}^{(b')}\right)~~\forall~m = 1, \dots, M; b = 1, \dots, B$ 
\State $P^{(b)} \gets \min_{m = 1, \cdots, M} P_{m}^{(b)}~~\forall~b = 1, \dots, B$
\State $P_{AdaMant} \gets \frac{1}{B + 1} \sum_{b = 0}^B I\left(P^{(0)} \leq P^{(b)}\right)$
\EndProcedure
\end{algorithmic}
\label{alg:alg1}
\end{algorithm}

\noindent \textbf{Remark} Algorithm \ref{alg:alg1} computes the AdaMant $P$-value via a permutation procedure.  This method can be computationally expensive, but is likely feasible in practice, as discussed below (Section \ref{computational-complexity}).  It is possible to compute the asymptotic reference distribution of the AdaMant test statistic using a moment-matching normal approximation and results on the order statistics of normal random variables \parencite{afonja1972moments}.  Details of this derivation are given in \ref{mantel_asymptotic}.  However, this normal approximation for a mixture of chi-squared random variables is not accurate for small samples \parencite{zhang2005approximate}, and so the permutation procedure is generally recommended.\\ 

\begin{figure}[ht]
\centering
\includegraphics[scale=0.37]{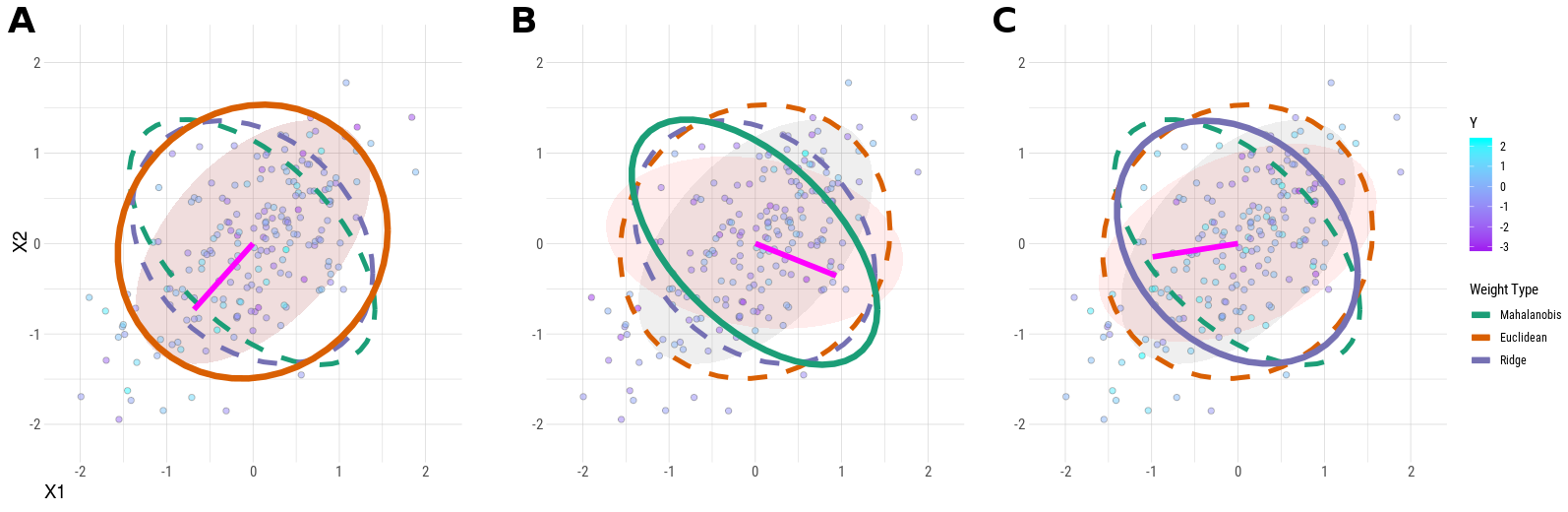}
\caption{Illustration of AdaMant for $Y$ generated from covariates $X$ rotated by angle $\theta$.  Gray shaded ellipses show the true covariance of $X$; pink shaded ellipses show the rotated covariances used to generate $Y$; magenta line segments show the direction of association estimated from a fixed effects model.  The solid lined ellipses correspond to the weights chosen by AdaMant; dashed ellipses are other weightings considered.  In each case, AdaMant selects the weighting that most emphasizes the empirical direction of association of $X$ and $Y$. (A) $\theta = 0$; (B) $\theta = 0.12 \pi$; (C) $\theta = 0.3 \pi$.}
\label{fig:adamant_three_plots}
\end{figure}

Let us now illustrate the utility of AdaMant through an example varying the relationship of the PCs of $X$ with $Y$ generated from a variance components model.  Suppose that $X$ is a $100 \times 2$ matrix of covariates drawn iid from a standard normal distribution, and $Y \sim N(0, \sigma^2 K^{(\theta)} + \sigma^2_{\varepsilon})$, with $K^{(\theta)} = U\Theta \Delta^2 \Theta^T U^T$ for $\Theta$ the $2 \times 2$ rotation matrix for angle $\theta$.  Figure \ref{fig:adamant_three_plots} shows the results of applying AdaMant with weights for the Euclidean, Mahalanobis, and ridge kernels ($\lambda = 10$) for data generated with $\theta = 0, 0.12 \pi, 0.3\pi$.  In each case, AdaMant selects the weighting that most emphasizes the empirical direction of association of $X$ and $Y$. This clearly demonstrates the potential advantage of AdaMant if the direction of association of $X$ and $Y$ is different than the first PC of $X$.  In this case, AdaMant will likely be a more powerful test than methods such as the RV test, which corresponds to using only the Euclidean weighting.

\subsection{Computational Complexity}
\label{computational-complexity}

If the feature space is very high-dimensional or if $n$ is large, a straightforward implementation of Algorithm 1 may be computationally impractical.  However, when only ridge kernels with varying values of $\lambda$ are included in AdaMant, there are two approaches that can be used to greatly reduce the computational cost.

The first approach utilizes the SVD $X = UDV^T$.  The computational complexity needed for finding the SVD for $X$ is $O(np^2)$. Once the SVD is computed, we can compute $Z=U^TY$ and $S_\lambda=\sum_{i=1}^r \frac{\eta_i}{\lambda+\eta_i}Z_i^2$, which has a total complexity of $O(nr)$. Note that when $p\gg n$, the rank $r$ is often the same as $n$; as a result, the cost needed for calculating $S_\lambda$ is $O(nr)=O(n^2)$. Calculating the test statistics for $B$ permutations requires $O(Bn^2)$, for a total computational complexity of $O(np^2+ Bn^2)$. 

Alternatively, when $p$ is very large relative to $n$, we can use the identity $X(X^TX + \lambda I)^{-1} = (XX^T + \lambda I)^{-1}X,$ so that the dimension of the matrix to be inverted is lowered to $n\times n$, rather than $p\times p$.  From this identity, $K_\lambda$ can be rewritten as 
$K_\lambda = (XX^T + \lambda I_n)^{-1}XX^T.$  Note that calculating $K_\lambda$ involves multiplying the $n\times p$ matrix $X$ and the $p\times n$ matrix $X^T$, multiplying two $n\times n$ matrices, and inverting an $n\times n$ matrix. When $p\gg n$, the computation cost is dominated by calculating $XX^T$, which has a complexity of $O(n^2p)$. The Mantel test statistic can be calculated as
$$T_\lambda = \text{tr}(YY^T K_\lambda)=Y^T K_\lambda Y,$$
which has a complexity of $O(n^2)$. With $B$ permutations, the total computational complexity is $O(n^2p + Bn^2)$, which is less than the required computational complexity using SVD. Thus, switching from the feature space to the subject space (i.e., from a $p\times p$ similarity matrix of the features to an $n\times n$ similarity matrix of the subjects), has a computational advantage.

\subsection{The Ridge Penalty and Variance Explained}
\label{ridge-penalty}

A main advantage of AdaMant is to allow for simultaneous testing over a set of tuning parameter values while incurring substantially less loss of power compared to multiple testing adjustment methods such as Bonferroni.  Although it does not impose an overly conservative adjustment for multiple testing, the power of AdaMant does decrease as the number of metrics considered increases.  Conversely, the test results are highly sensitive to the choice of parameters to test, thus one must take some care in the selection of the included parameters, balancing the desire to use a wide range of parameter values, with the gains of using a small set of parameters.  When only ridge kernels are included in AdaMant, previous results on the role of the ridge penalty term in predictive modeling can help with the identification of a reasonable set of values to test. 

Specifically, it has been shown that when the ridge penalty $\lambda$ is chosen to be the noise-to-signal ratio $\sigma_{\varepsilon}^2/\sigma^2 = (1 - h^2) / h^2$, the resulting shrinkage estimator $\hat\beta_\lambda$ is identical to the best linear unbiased predictor (BLUP) for the random effects model, and moreover, for a new observation with unknown response the predictions using the ridge with and random effects models are the same for $\lambda = \sigma_{\varepsilon}^2/\sigma^2$ \parencite{de2013prediction}.  Consequently, a reasonable set of penalty terms can usually be determined in practice by positing \textit{a priori} a likely range for the noise to signal ratio or a related quantity such as the proportion of variance explained.

The scientific interpretation and range of plausible values of $h^2$ will depend on the specific modalities of $X$ and $Y$. For instance, a genome-wide heritability of $h^2 > 0.5$ is very high (and would likely have already been discovered by previous studies), whereas a heritability of $h^2 < 0.1$ is probably not scientifically interesting.  As a point of reference, most estimated heritability in the UK Biobank data is between 0.1 to 0.4  \parencite{ge2017phenome}.

\begin{center}
\section{SIMULATIONS}\label{simulations}
\end{center}

\subsection{Univariate Simulations}
\begin{figure}[ht]
\centering
\includegraphics[scale = 0.8]{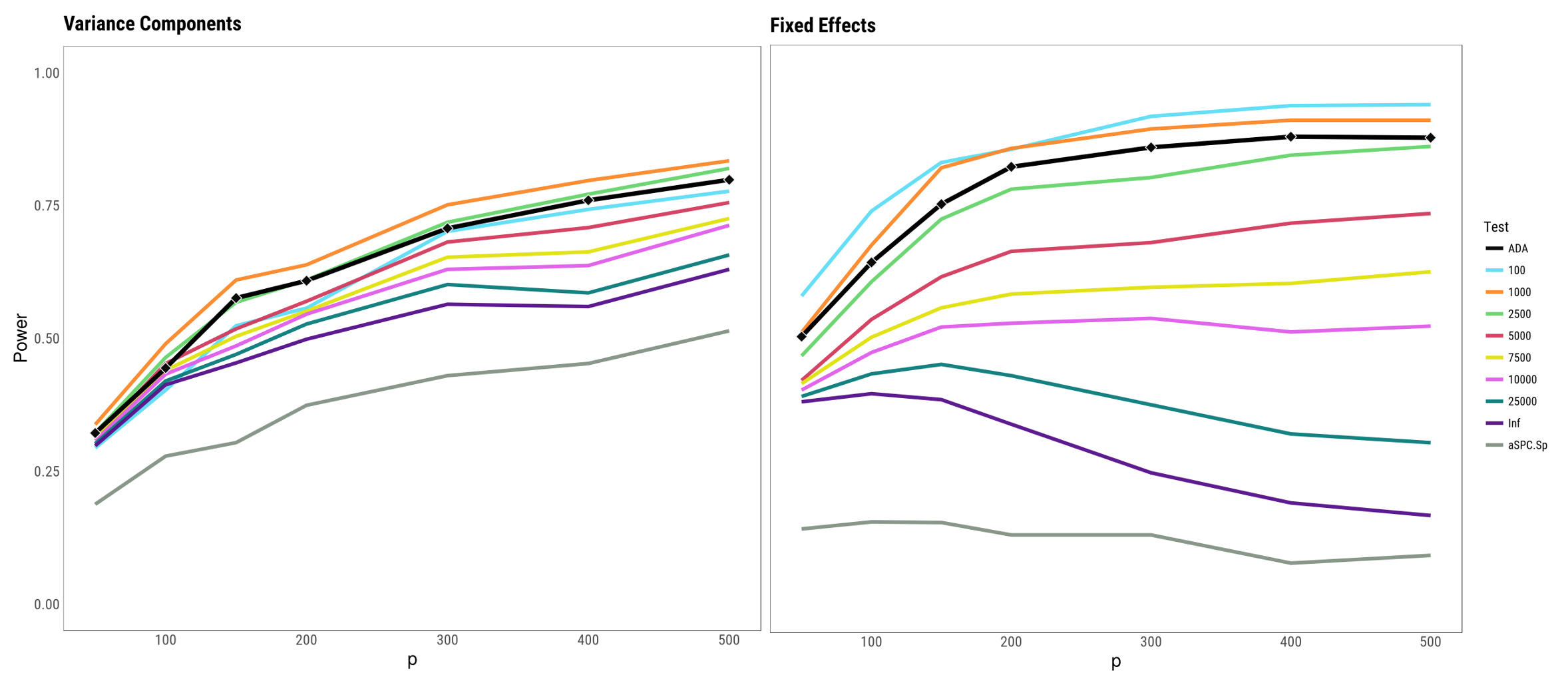}
\caption{Simulation results of the adaptive Mantel test.  Each setting used $n = 200$ observations, $\lambda \in \{100, 10^3, 2.5 \times 10^3, 5 \times 10^3, 7.5 \times 10^3, 10^4, 2.5 \times 10^4, \infty\}$, 500 test permutations, and 500 reps per value of $p$.  Error variance was fixed at $\sigma_{\varepsilon}^2 = 1.$ The black curve is the adaptive Mantel power; the other curves are the power for the simple Mantel test with the ridge kernel with indicated penalty term, and the aSPC test.  The dCov and RV-coefficient tests perform nearly identical to the $\lambda = \infty$ test.  \textbf{(i)} Results for data generated from the variance components model Eq. \ref{eqn:random} with constant effect size $\sigma_b = 0.035$ for each included feature. \textbf{(ii)} Power for data generated from a fixed effects model with $\beta_j = (-1)^j 0.05$.}
\label{fig:sims}
\end{figure}

To assess the performance of AdaMant for univariate phenotypes, the simulated response $Y$ was generated from either the variance components model (Eq. \ref{eqn:random}) or the fixed effects model (Eq. \ref{eqn:fixed}), with $n = 200$ observations, number of covariates $p$ ranging from 100 to 500, and $\sigma^2_e = 1$ fixed.  The design matrix $X$ was generated from $n$ draws from a $p$-variate normal distribution $N(0, \Sigma_X)$, where $\Sigma_X$ is chosen to have a compound symmetric structure with $\rho = 0.1$.  AdaMant was applied using the \textit{adamant} R-package \if1\blind{\parencite{pluta2018}} \fi  with $H = YY^T$ and $K_{\lambda}$ for $\lambda \in \{100, 10^3, 2.5 \times 10^3, 5 \times 10^3, 7.5 \times 10^3, 10^4, 2.5 \times 10^4, \infty\}$ (where $\infty$ corresponds to the Euclidean inner product similarity measure), with 500 permutations. For each setting, 500 simulations were run. 
Figure \ref{fig:sims} shows the simulation results for data generated from the variance components model with $\sigma^2_b = 0.035^2$, and the fixed effects model with $\beta_j = (-1)^j 0.05, j = 1, \dots, p$.

The simulation results indicate that the power of AdaMant is competitive with the best of the single-parameter MT for the values considered.  For the variance components setting, $\lambda = 1000$ exhibited the highest power for the ridge MT.  As anticipated, the power of the ridge MT exhibits unimodal behavior, such that the power increases to its peak for $\lambda = 1000$, and decreases as $\lambda \to \infty$.  We note that even though the variance components model is the true data generating mechanism, the test power is substantially increased over the variance components score test through the use of the ridge kernels.


When the response data is generated from the fixed effects model, the simulation results show that using a penalized ridge kernel produces much higher power than competing tests, with $\lambda = 100$ giving the highest power across all values of $p$.  These results suggest that the variance components test is severely underpowered when the relationship of $X$ and $Y$ is strongest along principal components of $X$ with low variance.  Moreover, the RV coefficient, dCov, and aSPC tests perform equally poorly in this setting.

Additional simulation results for sparse effects (setting the effect size of some features of $X$ to zero) are provided in the online supplement \ref{S-simulations}.

\subsection{Simulated EEG Data}



\begin{figure}[ht]
\centering
\includegraphics[scale = 0.6]{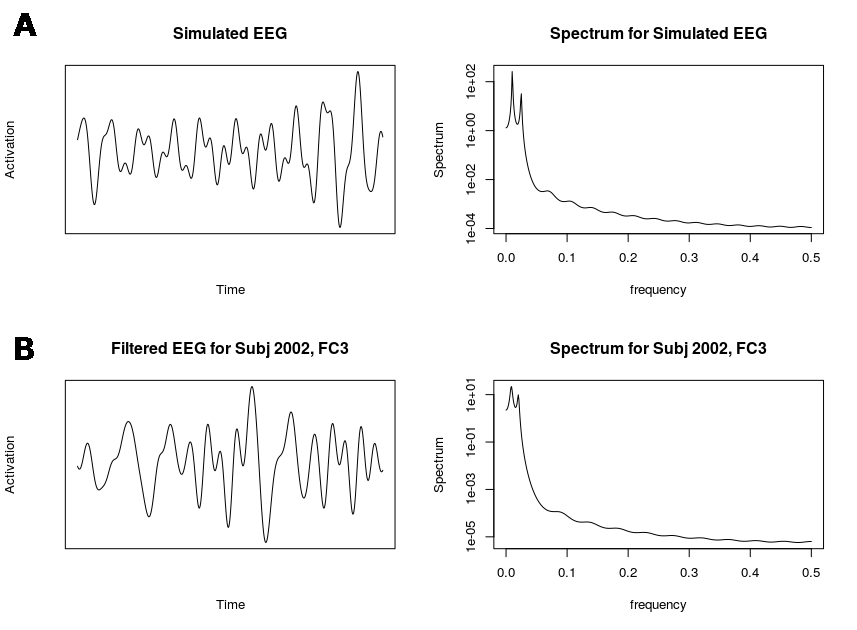}
\caption{A comparison of the simulated and real EEG time series shows the simulated series exhibit similar spectral characteristics as the real data after applying the same Butterworth bandpass filter (5 Hz - 50 Hz) (\textbf{Row A}) Filtered simulated EEG time series and corresponding spectrum. (\textbf{Row B}) Filtered EEG for subject 2002, channel FC4, and corresponding spectrum.}
\label{fig:simulated_eeg}
\end{figure}

Since we are primarily interested in applying AdaMant to imaging genetics data, we also consider the test performance on simulated data with structure similar to realistic EEG and SNP data (see Figure \ref{fig:simulated_eeg}).  To generate the genetic data, observations were placed into one of two balanced groups to form two groups of genetically similar observations.  For group $k, k = 1, 2$, a group-wide SNP vector $X^{(k)}$ of length 300 is generated by iid sampling from a discrete uniform distribution over $\{-1, 0, 1\}$; for subject $i$ in group $k$, a subject-specfic error vector $\varepsilon^{(k)}_{i} \sim N(0, I_{300})$ are added to the group SNP vector to get the subject SNP vector $X^{(k)}_i = X^{(k)} + \varepsilon^{(k)}_i$.  

To generate data with spectral characteristics similar to real EEG data, for each observation 20 time series of length 1000 were generated (for 20 ``channels'') as a weighted mixture of two independent AR(2) processes with spectral peaks at 5.12 Hz and 12.8 Hz and second-order AR-coefficient fixed at -1, resulting in a time series with two peaks at the chosen frequencies, with the magnitude of the peaks determined by the mixing weights.  The mixing weights for a random selection of 10 of the channels (common across all subjects) are drawn iid from $N(0, 1)$, so that no genetic effect exists for these channels.  To link the genetic features to the remaining channels, the $200 \times 10 \times 2$ array $W$ of AR(2) mixing weights was generated from a multivariate variance components model, such that for channel $j$ and AR(2) basis $m$, the subject weight vector is generated from $W_{jm} \sim N(0, \sigma^2_g YY^T)$, where $\sigma^2_g$ is varied across simulations to control the strength of genetic association.  The pair of weights for each subject and channel are then each squared and divided so that the weights sum to one.  From the resulting EEG data, the connectivity matrix for each observation is calculated from either the pairwise coherence for all pairs of channels.  

\begin{figure}[ht]
\centering
\includegraphics[scale = 0.4]{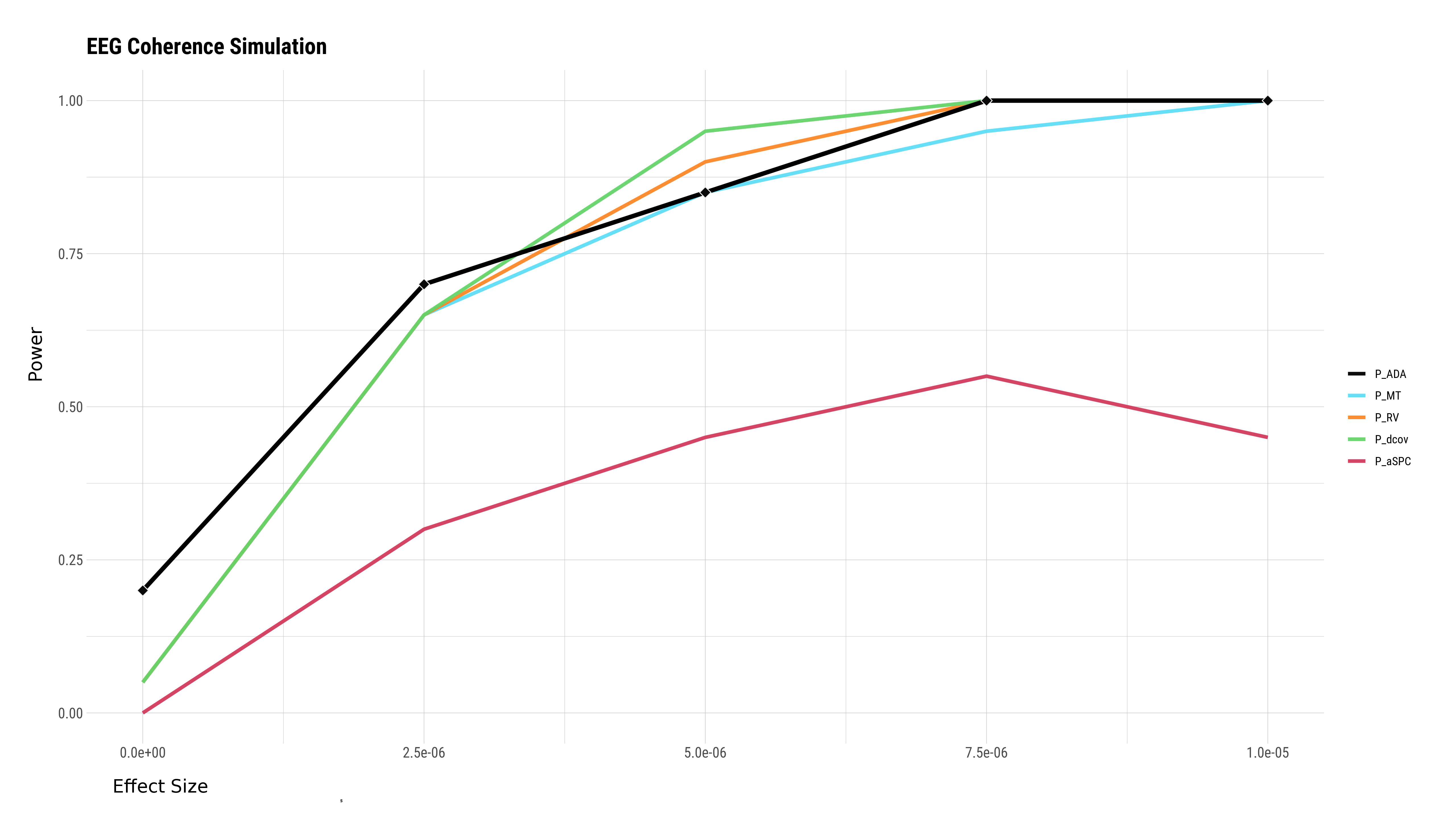}
\caption{Power plot of association testing methods for simulated 20-channel EEG data with $n = 200$ observations, $p = 300$ genetic features with aggregate effect size per channel per frequency equal to $\sigma^2$.  Subject similarity of coherence }
\label{fig:mv_sim}
\end{figure}

In applying AdaMant, similarity of observations in the EEG domain is then computed from the ridge kernel applied to the vectorized upper triangles of the observations' theta coherence matrices (resulting in 190 features), with candidate penalty terms $\Lambda_Y \in \{10, 100, \infty\}$.  Similarity in the genetics domain is computed from the ridge kernel applied to the observations' SNP vectors, with candidate penalty terms $\Lambda_X = \{10, 100, \infty\}$.  500 simulations were run for effect sizes equal to $\sigma = \{0, 2.5 \times 10^{-6}, 5 \times 10^{-6}, 7.5 \times 10^{-6}, 1 \times 10^{-5}\}$.  The simulation results show (Figure \ref{fig:mv_sim}) that AdaMant again has power competitive with other association testing methods in this setting.

\begin{center}
\section{APPLICATION TO IMAGING GENETICS DATA}
\label{application}
\end{center}

We next consider data from 350 healthy college students \if1\blind{ from Beijing Normal University (BNU)} \fi who participated in a visual working memory task conducted by our co-author, during which 64-channel EEG was recorded at 1 kHz.  These data are part of a broader set of imaging, genetics, and behavioral data collected with the goal of identifying neurological features of brain connectivity (measured via EEG) that are significantly associated with both genetic features and cognitive performance.  In the present study, we focus on testing the association of EEG coherence with a set of genome-wide SNPs, a set of candidate SNPs identified in a meta-analysis of educational attainment (EA), and a set of candidate SNPs previously implicated as potential factors in Alzheimer's Disease (AD).  

The coherence of two signals is a measure of the oscillatory correspondence (or spectral similarity) of two signals within a specified frequency band.   An outline of computing the coherence of two time series using the discrete Fourier transform is given in \ref{coherence}; a more comprehensive treatment is provided in \citeauthor{shumway2011time} (\citeyear{shumway2011time}), and \citeauthor{ombao2008evolutionary} (\citeyear{ombao2008evolutionary}).  EEG coherence has been implicated in many cognitive processes and neurological diseases, for instance, in associative learning \parencite{fiecas2016modeling}, as a predictor of ability in certain linguistic and visuo-spatial tasks \parencite{kang2017difference}, and in Alzheimer's disease \parencite{engels2015declining, vecchio2018sustainable} and schizophrenia \parencite{griesmayr2014eeg}.

\begin{figure}[ht]
\centering
\includegraphics[scale = 0.45]{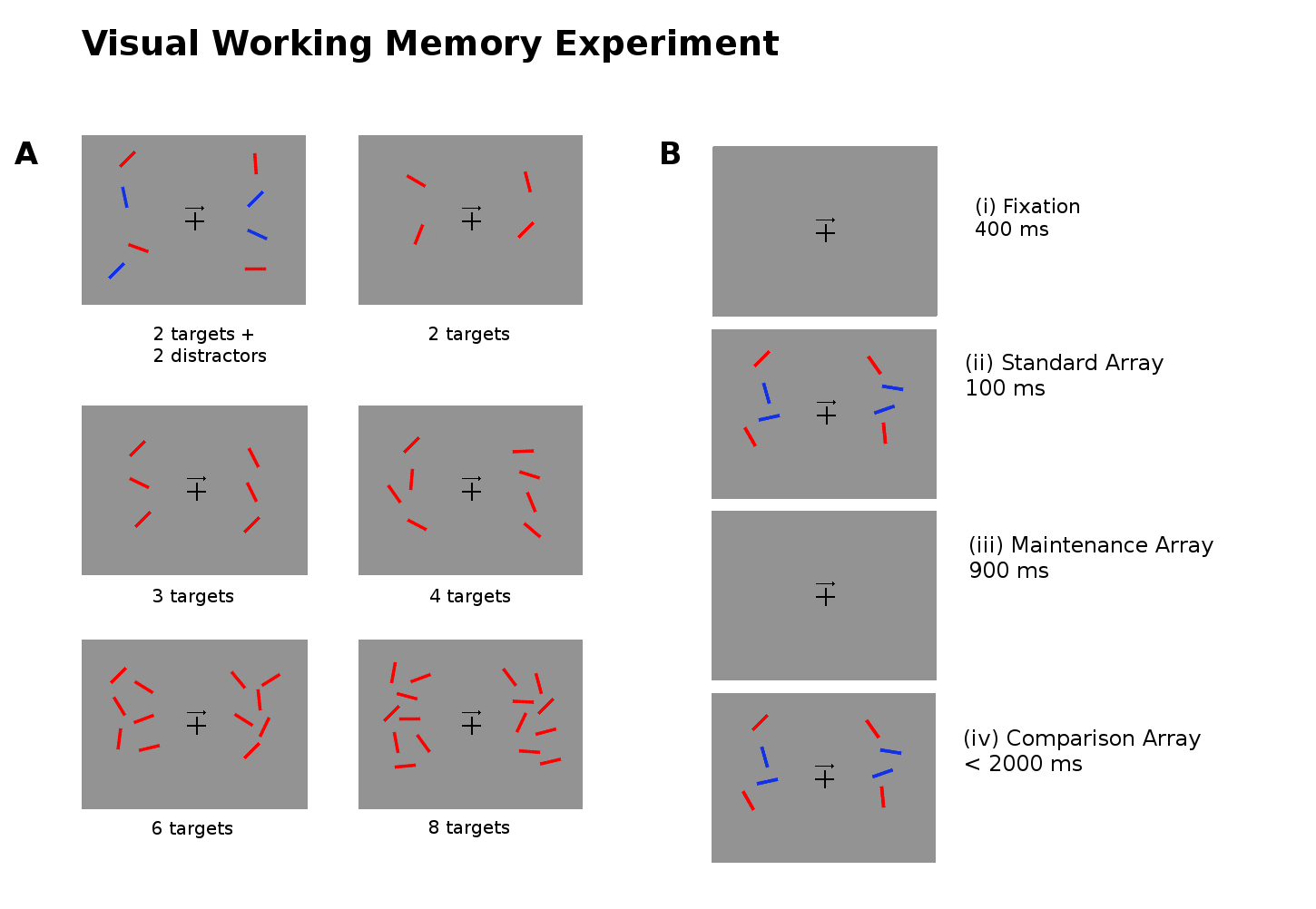}
\caption{(A) The experimental task required subjects to remember the positioning of the red bars (the targets) on the left or right side of the image as indicated by the arrow in the center of the image.  The six images show examples of the test image for different numbers of targets and distractors. (B) For each trial of the experiment, subjects are shown (i) a fixation image ($400 \pm 200$ ms); (ii) the standard array to be memorized (100 ms); (iii) the maintenance image (900 ms); (iv) the comparison array, at which point subjects are asked to respond within 2000 ms if the targets in the comparison arrays \textit{match} or \textit{mismatch} the standard array.}
\label{fig:vwm_experiment_design_new}
\end{figure}

The experimental task is depicted in Figure \ref{fig:vwm_experiment_design_new}. Subjects were asked to remember the positioning of the red bars (the targets) on the left or right side of the image as indicated by the arrow in the center of the image.  Six experimental conditions were used for 2, 3, 4, 6, and 8 targets, and for 2 targets + 2 blue distractors.  Each trial consists of the presentation of four images in sequence, with a total duration approximately three seconds, structured as follows: (i) \textit{fixation} image for $400 \pm 200$ ms; (ii) the \textit{standard array} to be memorized (100 ms); (iii) the blank \textit{maintenance} image (900 ms); (iv) \textit{comparison array} is shown, and subjects are prompted to respond within 2000 ms if the targets in the comparison array match or mismatch the standard array.  The total duration of the experiment was 5-10 minutes for each subject.  EEG data was downsampled to 256 Hz, channels with corrupted signals removed, and a band pass filter from 1 -- 40 Hz was applied. Independent components analysis was then used to remove physiological artifacts such as eyeblinks and other muscle movement \parencite{delorme2004eeglab, makeig2004mining}.  Lastly, all subjects were manually inspected to remove any remaining artifacts. The coherence between two EEG channels at a particular frequency $\omega$ is a measure of the oscillatory concordance of the the two signals at $\omega$.  Details of calculating the coherence are given in \ref{coherence}.  The pairwise coherence for $q$ EEG channels is a $q \times q$ symmetric matrix, from which we extract the upper triangle and vectorize to form the $n \times \genfrac{(}{)}{0pt}{}{q}{2}$ matrix $Y$.  

Channel topography and coordinates are available from the BioSemi website \parencite{biosemi}.  For our analysis, we consider coherence restricted to the 25 most anterior channels (``Frontal''), and fronto-parietal (FP) coherence from 30 channels covering the frontal and parietal lobes. This results in 300 distinct features for the 25 frontal channels, and 435 features for the fronto-parietal channels.  Figure \ref{fig:biosemi_channel_groups} shows the location of the channels for the frontal and fronto-parietal groups.

\subsection{Heritability of EEG Coherence}
Genotype data from approximately $5 \times 10^5$ single nucleotide polymorphisms (SNPs) for all subjects was measured, of which 484496 autosomal SNPs meeting quality control thresholds ($MAF > 1\%, HWE < 0.001$) were included for heritability testing.  Data processing and heritability testing was conducted with PLINK v1.90b4.4 and GCTA 1.91.5 beta2.  Let $X$ be the column-standardized SNP data matrix.

To determine whether the heritability of EEG coherence in the theta, alpha, and beta bands is significantly greater than zero, AdaMant was performed with $K = XX^T$ (which is proportional to the usual genetic relationship matrix), $H = Y(Y^TY + \lambda I)^{-1}Y^T$ for $\lambda \in \{10, 100, 1000, \infty\}$, using 5000 permutations.  

The most significantly heritable coherence was found to be the theta band (4 -- 8 Hz) for the frontal channels, with $P = 0.0458$.  Other bands tested were not found to be significantly heritable, additional results can be found in  Table \ref{tab:gwas_mantel}.  The estimated mean heritability of theta coherence for the frontal channels is $\hat h^2 = 0.165$, although there is large variance in this estimate due to the somewhat small sample size.

\subsection{Association of EEG Coherence and SNPs related to Educational Attainment}

Following the scientific aims of the \if1\blind{BNU} \fi \if0\blind{present} \fi imaging genetics study, it is of interest to determine whether genetic markers linked to educational attainment (EA) are significantly associated with EEG coherence during the visual working memory experiment.  SNPs were selected from the meta-analysis of educational attainment by \citeauthor{okbay2016genome} (\citeyear{okbay2016genome}), which identified over $3 \times 10^5$ SNPs found to be significantly associated with EA at a level of 0.01.  These SNPs were also found to show significantly elevated expression specifically in developing neural tissue, thus it is plausible they may be significantly associated with other aspects of cognitive development and performance.  

Testing the association of the top 100 EA SNPs with EEG coherence, we found that similarities from beta and gamma band coherence are significantly associated with genetic similarity measured from the top 100 SNPs from the EA gene set for the Frontal and Fronto-parietal coherence.  The beta band was the most significantly associated, with $P = 0.0104$ for the frontal channels and $P = 0.013$ for the FP channels.  

\subsection{Testing Association of EEG Coherence and SNPs Related to Alzheimer's Disease}

For a more targeted test of genetic influence on EEG coherence, a group of 11 SNPs that have been identified as potentially related to Alzheimer's Disease were selected in order to determine if these SNPs are related to working memory brain function in healthy individuals.  AdaMant test was performed with $\lambda \in \{0.5, 1, 5, 10, 100, 1000, \infty\}$ for ridge kernel similarity of the coherence data, and using 5000 permutations.  Genetic similarity of subjects was calculated as the $L_2$ inner product of the standardized SNP data for all tests.  For the alpha band, AdaMant $P$-values were $P = 0.065$ and $P = 0.381$ for all channels and frontal channels respectively.  Results for the theta band were $P = 0.416$ and $P = 0.085$ for all 64 channels and frontal channels respectively.  Since the adaptive Mantel test was used, these $P$-values already take into account testing across multiple $\lambda$. In the case of the alpha band, the test results suggest that coherence involving channels outside of the selected frontal channels may be associated with genetic similarity determined by the 11 AD SNPs, whereas for the theta band, the SNP association appears stronger when considering only the frontal channels.

Additional results on the specific channels and SNPs that are most associated, and discussion of the scientific relevance is provided in the online supplement \ref{additional-results}.

\section{Discussion}\label{discussion}

We have here developed the adaptive Mantel test as a metric-based testing procedure, which reduces the difficulty of kernel parameter selection by simultaneously testing across a set of candidate parameters and automatically selecting the parameter yielding the most significant test from the candidate set.  As a tool for high-dimensional inference, AdaMant improves over competing association tests by data-adaptively selecting the kernel that yields the most significant association of $X$ and $Y$, without the need to explicitly adjust the calculated $P$-value.  Moreover, AdaMant is simple to implement, yet can be usefully applied for any data modality that admits a meaningful measure of similarity.  As part of the present contributions, and to facilitate the use of AdaMant by other researchers, an R-package providing a basic implementation of AdaMant is available on the author's website \if1\blind{\parencite{pluta2018}}\fi.

To motivate and justify AdaMant, we have developed a unified framework of linear model score tests that links the score tests of the variance components, fixed effects, and ridge regression models as a single class of tests indexed by the penalty term $\lambda \in [0, \infty]$, with the fixed effects and variance components tests corresponding respectively to the extreme values of $\lambda$ in this range.  Through this framework, we have shown that linear model score tests are equivalent to testing the correlation of subject-subject similarities in each of the feature sets, that is, the linear model score tests are equivalent to the Mantel test for specific choices of metrics.   

From the forms of the tests given in Section \ref{score_tests}, and illustrated in Figure \ref{fig:adamant_three_plots}, we observe that the variance components test is preferred when the relationship of $X$ and $Y$ is most significant along the principal directions of $X$, whereas the fixed effects model is preferred when $X$ and $Y$ are significantly related along principal directions of $X$ with small variance. Since the fixed effects test cannot be applied when $n < p$, the ridge regression score test is an important generalization of the variance components and fixed effects tests, allowing one to recover the advantages of the fixed effects test in settings with $p > n$.

The present paper has primarily focused on the properties of AdaMant when only ridge kernel similarities are used, but this class of weights may not be optimal.  For instance, in Figure \ref{fig:adamant_three_plots} (B-C), weights that are concentrated along the true direction of association will produce a more powerful test than the ridge kernel weights.  Considering a wider class of similarity metrics, such as rotations of the Mahalanobis metric or non-linear kernel functions, is an important direction of future work for further improving the power and efficiency of high-dimensional testing methods.

AdaMant, along with many other currently used association testing methods, are powered for only a specific functional association of the two feature sets, e.g. linear or quadratic.  An important exception to this is the dCov test with the Euclidean distance, which detects any type of statistical dependence of the two feature sets \parencite{szekely2007measuring}.  As an alternative to this, AdaMant with the Gaussian radial basis kernel and a suitable selection of tuning parameters may be able to detect a wide variety of associations (although not all).  Building on kernel-based methods for estimating heritability \parencite{liu2007semiparametric}, future work will consider extensions of AdaMant to a more general class of similarity measures, and investigate methods for improving testing power while selecting across a large number of tuning parameters.

\newpage
\printbibliography
\newpage
\newpage
\appendix
\makeatletter
\makeatother
\begin{center}
\section{APPENDIX}
\end{center}

\subsection{Proofs of Propositions}
\subsubsection{Proposition 1}
\label{prop1}

Let $X$ be a column-centered $n \times p$ matrix with $\text{rank}(r)$, SVD $X = UDV^T$, and squared singular values $\eta_j, j = 1, \cdots, r$.  Let $Y \sim N(0, \sigma^2 \Sigma_A + \sigma^2_{\varepsilon} I_n)$ be an $n \times 1$ response vector, where $\Sigma_A$ is an $n \times n$ semi-positive definite covariance matrix that depends on $X$.  Define $Z = U^TY$, the $r \times 1$ vector of principal correlations, and define $T_F, T_V$, and $T_{\lambda}$ by Equations \ref{eqn:score_F}, \ref{eqn:score_V}, and \ref{eqn:score_lambda}.

\begin{enumerate}
\item \textbf{Connections between linear model score tests.} The score tests for the fixed effects, variance components, and ridge regression models form a single class of tests parameterized by the ridge tuning parameter $\lambda \in [0, \infty]$. $T_{\lambda}$ is testing equivalent to the variance components score test as $\lambda \to \infty$, $\lim_{\lambda \to \infty} T_{\lambda} \asymp T_V.$ Assuming $\text{rank}(X) = p$, $T_{\lambda = 0}$ is testing equivalent to the fixed effects score statistic, $T_0 \asymp T_F$.  

\item \textbf{Score test reference distributions.}
Let $\chi_{1, j}^2, j = 1, \dots, r$ be iid chi-squared random variables with one degree of freedom.  Let $\chi_p^2$ be a chi-squared random variable with $p$ degrees of freedom.
\begin{enumerate}
\item $T_{\lambda} \asymp \sum_{j=1}^{r}\frac{\lambda\eta_j}{\lambda+\eta_j} Z_j^2  ~\overset{H_0} \sim ~ \sum_{j=1}^{r}\frac{\lambda\eta_j}{\lambda+\eta_j} \chi^2_{1, j}.$
\item $T_F \asymp \sum_{j = 1}^p Z_j^2 ~\overset{H_0} \sim ~ \chi_p^2.$  
\item $T_V \asymp \sum_{j = 1}^r \eta_j Z_j^2 ~\overset{H_0} \sim ~ \sum_{j = 1}^r \eta_j \chi_{1, j}^2.$
\end{enumerate}
\end{enumerate}

\noindent \textbf{Proof}

\begin{enumerate}
    \item 
    \begin{align*}
        \lim_{\lambda \to \infty} T_{\lambda} &\asymp \lim_{\lambda \to \infty} \sum_{j = 1}^r \frac{\lambda \eta_j}{\lambda + \eta_j} Z_j^2\\
        &= \lim_{\lambda \to \infty} \sum_{j = 1}^r \frac{\eta_j}{1 + \eta_j / \lambda} Z_j^2\\
        &= \sum_{j = 1}^r \eta_j Z_j^2 \asymp T_V
    \end{align*}
    
    \item Under the null hypothesis of $H_0: \sigma^2 = 0$, the reference distribution is $Z = U^T Y \overset{H_0}\sim N(0, I_r)$.  Since each $Z_j$ is iid standard normal, $Z_j^2 \sim \chi^2_{1, j}$ for $chi^2_{1, j}$ iid copies of a chi-squared random variable with one degree of freedom.
\end{enumerate}

\subsubsection{Proposition 2}
\label{prop2}

Let $X$ be an $n \times r$ column centered matrix of covariates with $\text{rank}(X) = r$ and singular value decomposition $X = UDV^T$, with singular values $d_j, j = 1, \cdots, r$. For convenience, we define $\eta_j=d_j^2, j=1,\cdots, r$. Suppose $Y$ be an $n \times 1$ response vector, and let \(H=YY^T\) and $Z = U^TY$.  Assume that $Y$ is standardized so that $\text{tr}(H) = n$ and $\text{tr}(H^2) = n^2$.

\begin{enumerate}
	\item Fixed Effects 
    	\[R(H, K_F) = \frac{\sum_{j=1}^r Z_j^2}{\sqrt{np}} = \frac{1}{\sqrt{p}}R^2(X,Y).\]
        For simplicity, here we assume $p=r$. Note that when $X^TX$ is not full rank, one can replace $(X^TX)^{-1}$ with $(X^TX)^-$and the similarity matrix $X(X^TX)^{-1}X^T$ is invariant to the choice of a generalized inverse. 
	\item Random Effects 
    	\[R(H, K_V) = \frac{\sum_{j=1}^r\eta_i Z_j^2}{n\sqrt{\sum_{j=1}^r\eta_j^2}}\]
    \item Ridge Regression
    	\[R(H, K_{\lambda}) = 
        \frac{\sum_{j=1}^{r} Z_j^2 \frac{\eta_j}{\eta_j+\lambda}}{n\sqrt{\sum_{j=1}^{r} \left(\frac{\eta_j}{\eta_j+\lambda}\right)^2}} \]
\end{enumerate}

\noindent \textbf{Proof}

\noindent (1) The matrix $K_F$ is the standard projection matrix (the ``hat matrix'') from linear regression, giving the identity $\text{tr}(KK) = \text{tr}(K) = \text{rank} (K) = p = r.$  The result follows directly from the definition of $R(H, K)$ and $R^2(X, Y)$

\begin{align*}
R(H,K) &= \frac{\text{tr}(HK)}{\sqrt{\text{tr}(HH)\text{tr}(KK)}} = \frac{Y^TX(X^TX)^{-1}X^TY}{\sqrt{p}Y^TY}=\frac{1}{\sqrt{p}}R^2(X,Y).
\end{align*}
If we replace \(X\) with its singular value decomposition, we have
\[R(H,K)=\frac{Y^TUDV^TV(D^TD)^{-1}V^TVD^TUY}{\sqrt{p}Y^TY}=\frac{\sum_{j=1}^{r}Z_j^2}{n\sqrt{p}}\]

\vspace{0.5cm}
\noindent (2) In a random-effect linear model, the variance-covariance structure of
the outcome vector is a function of the similarity \(K_V\) that is defined
using the dot product: \(K_V=XX^T\). Let \(H=YY^T\). We have
\begin{eqnarray*}
\text{tr}(HH)&=& = \left(\sum_{j=1}^r Z_j^2\right)^2 = n^2\\
\text{tr}(K_VK_V)&=& \text{tr}(XX^TXX^T)=\text{tr}(UDV^TVD^TU^TUDV^TVD^TU^T)=\sum_{j=1}^r\eta_j^2\\
\text{tr}(HK_V)&=& Y^T XX^T Y = Y^T U DD^T U^T Y = \sum_{j=1}^r \eta_j Z_j^2.
\end{eqnarray*}

\vspace{0.5cm}
\noindent (3) In a ridge regression, let $K_\lambda = X^TX + \lambda I_p$ and $H=YY^T$. We have
\begin{align*}
K_\lambda &= VD^TDV^T+\lambda VV^T=V(D^TD+\lambda I)V^T\\
\text{tr}(K_\lambda K_\lambda) &= \text{tr}(X(X^TX+\lambda I)^{-1} X^TX(X^TX+\lambda I)^{-1}X^T)\\
&= \sum_{j=1}^{r} \left(\frac{\eta_j}{\eta_j+\lambda}\right)^2.
\end{align*}
Let \(Z=U^TY=(Z_1,\cdots,Z_r)^T\), we have
\begin{eqnarray*}
Y^TY&=&\sum_{j=1}^r Z_j^2 = n\\
\text{tr}(HK_\lambda) &=& Y^TX(X^TX+\lambda I)^{-1} X^TY\\
&=& \sum_{j=1}^{r} \frac{\eta_j}{\eta_j+\lambda} Z_j^2 
\end{eqnarray*}

The correlation coefficient based on the similarity matrices is

\begin{eqnarray*}
R(H,K_\lambda) &=& \frac{\text{tr}(HK)}{\sqrt{\text{tr}(HH)\text{tr}(KK)}} =\frac{\sum_{j=1}^r  \frac{\eta_j}{\eta_j+\lambda}Z_j^2}
{n\sqrt{\sum_{j=1}^{r} \left(\frac{\eta_j}{\eta_j+\lambda}\right)^2 }}
 \end{eqnarray*}

\noindent (4) When \(\lambda=0\), we assume that \(p<n\). The result
follows immediately.

When \(\lambda>0\), let \(\delta=1/\lambda\) and express the correlation at \(\lambda\) in terms of \(\delta\):

\begin{eqnarray*}
R(H,K_{1/\delta}) &=& \frac{\sum_{j=1}^{r} \frac{\eta_j}{\eta_j+1/\delta} Z_j^2}{n\sqrt{\sum_{j=1}^{r} \left(\frac{\eta_j}{\eta_j+1/\delta}\right)^2 }}
\end{eqnarray*}

Using L'Hospital's rule, it is not difficult to see that
\[\lim_{\delta\rightarrow 0} R(H,K_{1/\delta})= \frac{\sum_{j=1}^{r} \eta_jZ_j^2}{n\sqrt{\sum_{j=1}^{r} \eta_j^2}}=R(H, K_V),\]
i.e.,
\[R(H,K_\infty)=\frac{\sum_{j=1}^r \eta_jZ_j^2}{n\sqrt{\sum_{j=1}^r \eta_j^2}}=R(H, K_V),\]

Note that \(R(H,K_\infty)\le R(H,K_0)\); in other words, the ridge
regression method shrinks the correlation. Another observation is that
when \(\eta_1=\eta_2=\cdots=\eta_r\), \(X\) has orthogonal
columns and all the correlations are identical.

\subsubsection{Proposition 3}
\label{prop3}

\noindent Assume $X$ is an $n \times p$ column-standardized, full row-rank matrix (which requires $p \geq n$), and
assume $Y \sim N(0, \sigma^2 I_q \otimes XX^T / p + \sigma^2_{\varepsilon}I_q \otimes I_n)$ is a $q$-variate column-standardized random vector. Define Gram matrices $G = XX^T / p$ and $H = YY^T$ (for which $\text{tr}(G) = n$ and $\text{tr}(H) = nq$ under the above assumptions).  

\begin{enumerate}
	\item The method of moments estimator $\hat h^2$ is
	
	$$\hat h^2 = \frac{\text{tr}(HG / q) - n}{\text{tr}(G^2) - n}.$$

	\item The expectation of the correlation of Gram matrices is
		$$\mathbb{E} R(H, G) = h^2\frac{(\text{tr}(G^2) - n)}{\sqrt{\text{tr}(H^2 / q)\text{tr}(G^2)}} + \frac{n}{\sqrt{\text{tr}(H^2 / q)\text{tr}(G^2)}}.$$
\end{enumerate}
\vspace{1cm}

\noindent \textbf{Proof}
\begin{enumerate}
\item 

First compute $\mathbb{E}~\text{tr}(HG)$ following standard results on the quadratic forms.

\begin{align*}
    \mathbb{E}~\text{tr}(HG) &= \text{tr}(G(\sigma^2 I_q \otimes G + I_q \otimes \sigma^2_{\varepsilon}I_n))\\
    &= q\sigma^2 \text{tr}(G^2) + nq\sigma^2_{\varepsilon}.
\end{align*}

\noindent Plugging in $\text{tr}(HG)$ and solving for $h^2$ gives the desired method of moments estimator

$$\hat h^2 = \frac{\text{tr}(HG / q) - n}{\text{tr}(G^2) - n}.$$

\item 
\begin{align*}
	\mathbb{E} R(H, G) &= \frac{1}{\sqrt{\text{tr}(H^2)\text{tr}(G^2)}} \mathbb{E}\text{tr}(HG)\\
	&= q\sigma^2 \text{tr}(G^2) + nq\sigma^2_{\varepsilon}\\
	&= h^2\left(\frac{\text{tr}(G^2) - n}{\sqrt{\text{tr}(H^2 / q)\text{tr}(G^2)}}\right) + \frac{n}{\sqrt{\text{tr}(H^2 / q)\text{tr}(G^2)}}
\end{align*}

\end{enumerate}

\noindent \textbf{Corollary} In the setting of Proposition 3, with $Y$ a univariate vector ($q = 1$) and for fixed $n$, $p$, and $\sigma^2$, with $p > n$, the expected correlation $\mathbb{E}R(H, G)$ is a function of $\text{tr}(G^2)$, with maximum and minimum dependent on the noise-to-signal ratio $(1 - h^2) / h^2$. 
	\begin{enumerate}
	    \item For $1 \leq (1 - h^2) / h^2 \leq n$, the expected correlation is bounded by
	    
	    $$2\sqrt{\frac{h^2(1 - h^2)}{n}} \leq \mathbb{E}R(H, G) \leq h^2\frac{n - 1}{n} + \frac{1}{n}.$$
	    
	    with upper and lower bounds attained by $\text{tr}(G^2) = n^2$ and $\text{tr}(G^2) = \frac{n(1 - h^2)}{h^2}$ respectively.
	    \item For $(1 - h^2) / h^2 \leq 1$, the expected correlation is bounded by
	    
	    $$\frac{1}{\sqrt{n}} \leq \mathbb{E}R(H, K) \leq h^2\frac{n - 1}{n} + \frac{1}{n},$$
	    
	    with upper and lower bounds attained at $\text{tr}(G^2) = n^2$ and $\text{tr}(G^2) = n$ respectively.
	    \item For $(1 - h^2) / h^2 \geq n$, the expected correlation is bounded by
	    
	    $$h^2\frac{n - 1}{n} + \frac{1}{n} \leq \mathbb{E}R(H, G) \leq \frac{1}{\sqrt{n}},$$
	    
	    with upper and lower bounds attained at $\text{tr}(G^2) = n$ and $\text{tr}(G^2) = n^2$ respectively.
	\end{enumerate}

\noindent \textbf{Proof}

Define the function $\rho(\xi | \sigma^2) = \frac{1}{n\sqrt{\xi}}\left(h^2(\xi - n) + n\right)$.  Here $\xi$ can be interpreted as a candidate value for $\text{tr}(G^2)$.  The derivative of $\rho$ is

$$\rho^{\prime}(\xi) = \frac{1}{2n}\xi^{-3/2}(h^2\xi + nh^2 - n).$$

\noindent It follows that $\xi^* = n\frac{1 - h^2}{h^2}$ gives $\rho'(\xi_*) = 0$. The signs of the derivative can be checked to show that this is in fact a minimum of $\rho$.  The maximum and minimum of $\rho$ attained on $n \leq \xi \leq n^2$ (the range of $\text{tr}(G^2)$ determined by $\text{tr}(G) = 1$) depends on the relationship of $\xi^*$ to the range of $\text{tr}(G^2)$, for which there are three cases.

\begin{enumerate}
	    \item When $1 \leq (1 - h^2) / h^2 \leq n$, $\xi^*$ is in the range of $\text{tr}(G^2)$, and so the minimum of $\rho$ is attained here.  Since $\rho$ is strictly increasing for $\xi > \xi^*$, the maximum of $\rho$ is attained at $\text{tr}(G^2) = n^2$.
	    
	    $$2\sqrt{\frac{h^2(1 - h^2)}{n}} \leq \mathbb{E}R(H, G) \leq h^2\frac{n - 1}{n} + \frac{1}{n}.$$
	    
	    with upper and lower bounds attained by $\text{tr}(G^2) = n^2$ and $\text{tr}(G^2) = \frac{n(1 - h^2)}{h^2}$, respectively.
	    
	    \item When $(1 - h^2) / h^2 \leq 1$, $\rho$ is strictly increasing in the range of $\text{tr}(G^2)$, so the maximum and minimum are attained are the right and left endpoints of the interval, respectively.
	    
	    \item When $(1 - h^2) / h^2 \geq n$, $\rho$ is strictly decreasing in the range of $\text{tr}(G^2)$, so the maximum and minimum are attained at the left and right endpoints of the interval, respectively.
	    
	    $$h^2\frac{n - 1}{n} + \frac{1}{n} \leq \mathbb{E}R(H, G) \leq \frac{1}{\sqrt{n}},$$
	    
	    with upper and lower bounds attained at $\text{tr}(G^2) = n$ and $\text{tr}(G^2) = n^2$ respectively.
\end{enumerate}

\subsection{Multivariate Ridge}
\label{multivariate_ridge}

Previous results have shown that ridge regression with tuning parameters for both $X$ and $Y$ can be effective for the simultaneous prediction of multiple outcome variables (\cite{brown1980adaptive,breiman1997predicting}), thus it may be useful to consider ridge kernel similarities for both $X$ and $Y$.  For two tuning parameters $\lambda_X, \lambda_Y > 0$ and Gram matrices $H_{\lambda_Y}=Y(Y^TY+\lambda_Y I_q)^{-1}Y^T \mbox{ , } K_{\lambda_X}=X(X^TX+\lambda_X I_p)^{-1}X^T$, the ridge kernel Mantel test statistic is $T_{\lambda_X,\lambda_Y} = \text{tr}(H_{\lambda_Y}K_{\lambda_X})$.  The corresponding augmented data model is

\begin{equation}
\tilde{Y}=\begin{pmatrix}Y\\ \mathbf{0}\\ \sqrt{\lambda_Y} I_q\end{pmatrix}, \tilde{X}=\begin{pmatrix}X\\ \sqrt{\lambda_X} I_p \\ \mathbf{0}\end{pmatrix}
\end{equation}

To derive the correspondence with the ridge kernel similarity measures on $X$ and $Y$, let $\tilde{H}=\tilde{Y} (\tilde{Y}^T\tilde{Y})^{-1}\tilde{Y}^T$, $\tilde{K}=\tilde{X} (\tilde{X}^T\tilde{X})^{-1}\tilde{X}^T$.  The least squares estimate (based on the augmented data) of $B$ is
$$\hat B=(\tilde X^T\tilde X)^{-1}\tilde X^T \tilde Y=(X^TX+\lambda_X I)^{-1}X^TY,$$
which only depends on the tuning parameter for $X$. To see the effect of the tuning parameter $\lambda_Y$, we examine the log-likelihood based on the augmented data,

\begin{equation}
\begin{split}
l \propto c &- (n+p+q)\log|\Sigma_{\varepsilon}| - \text{tr}[(Y-XB)\Sigma_{\varepsilon}^{-1}(Y-XB)^T]\\ 
&- \lambda_X\text{tr}[B\Sigma_{\varepsilon}^{-1}B^T]-\lambda_Y \text{tr}[\Sigma_{\varepsilon}^{-1}],
\label{eqn: neglik_augment}
\end{split}
\end{equation}

\noindent where $c$ is a constant that does not depend on any parameters.  

Another observation is that the negative likelihood function in Equation \ref{eqn: neglik_augment} is mathematically identical to the posterior distribution of $B$ and $\Sigma_e$ where the joint prior is a normal-inverse-Wishart distribution:
\begin{eqnarray}
Y|B,\Sigma_e &\sim& N_{n\times q}(XB, \Sigma_e, I_n)\\
p(B,\Sigma_e) &\propto&  |\Sigma_e|^{\frac{p+q}{2}}exp\{-\frac{\lambda_x}{2}tr[B\Sigma_e^{-1}B^T]-\frac{\lambda_y}{2} tr[\Sigma_e^{-1}]\}
\end{eqnarray}

\subsection{Adaptive Mantel Test Asymptotics}
\label{mantel_asymptotic}

Assume true model is the random effects $Y \sim N(0, \sigma^2 K_V + \sigma^2_{\varepsilon}I_n),$ where $K_V = XX^T$.  Assume the SVD of $X$ is $X = U D V^T$, with singular values $d_j$ and eigenvalues $\eta_j = d_j^2, j = 1, \cdots, p$.  Further assume that $X$ and $Y$ have been centered and scaled, so that $\text{tr}(XX^T) = np$, and $\text{tr}(X(X^TX)^{-1}) = p$.
  
\subsubsection{Random Effects Test}
  
Let $K_V = XX^T$.

\begin{align}
T_R &= \text{tr}(HK_V) = \sum_{j = 1}^p \eta_j Z_j^2\\
\mathbb{E}T_R &= \text{tr}(\sigma^2 K_V^2) + \sigma^2_{\varepsilon} np = \sigma^2 \sum \eta_j^2 + \sigma^2_{\varepsilon}np\\
T_R^* &= \frac{T_R - \sigma^2_{\varepsilon}np}{\sigma^2_{\varepsilon}\sqrt{2\sum \eta_j^2}} ~\overset{H_0}{\dot\sim}~ \mathcal{N}(0, 1)\\
\mathbb{E}T_R^* &= \frac{\mathbb{E}T_R - \sigma_{\varepsilon}^2np}{\sigma^2_{\varepsilon}\sqrt{2\sum \eta_j^2}} = \frac{\sigma^2}{\sigma^2_{\varepsilon}}\sqrt{\frac{1}{2}\sum\eta_j^2}
\end{align}

\subsubsection{Fixed Effects Test}

Let $K_F = X(X^TX)^{-1}X^T$.

\begin{align}
T_F &= \text{tr}(HK_F) = \sum_{j = 1}^p Z_j^2\\
\mathbb{E}T_F &= \text{tr}(\sigma^2 K_F K_V) + \sigma^2_{\varepsilon} p = \sigma^2 np + \sigma^2_{\varepsilon}p\\
T_F^* &= \frac{T_F - \sigma^2_{\varepsilon}p}{\sigma^2_{\varepsilon}\sqrt{2p}} ~\overset{H_0}{\dot\sim}~ \mathcal{N}(0, 1)\\
\mathbb{E}T_F^* &= \frac{\mathbb{E}T_F - \sigma_{\varepsilon}^2p}{\sigma^2_{\varepsilon}\sqrt{2p}} = \frac{\sigma^2}{\sigma^2_{\varepsilon}}\cdot n\sqrt{\frac{p}{2}}
\end{align}

\subsubsection{Ridge Kernel Test}

Let $K_{\lambda} = X(X^TX + \lambda I_p)^{-1}X^T$.

\begin{align}
T_{\lambda} &= \text{tr}(HK_{\lambda}) = \sum_{j = 1}^p \frac{\eta_j}{\eta_j + \lambda} Z_j^2\\
\mathbb{E}T_{\lambda} &= \text{tr}(\sigma^2 K_{\lambda} K_V) + \sigma^2_{\varepsilon} \sum \frac{\eta_j}{\eta_j + \lambda}\\
T_{\lambda}^* &= \frac{T_{\lambda} - \sigma^2_{\varepsilon}\sum \frac{\eta_j}{\eta_j + \lambda}}{\sigma^2_{\varepsilon}\sqrt{2\sum \left(\frac{\eta_j}{\eta_j + \lambda}\right)^2}} ~\overset{H_0}{\dot\sim}~ \mathcal{N}(0, 1)\\
\mathbb{E}T_{\lambda}^* &= \frac{\mathbb{E}T_{\lambda} - \sigma_{\varepsilon}^2 \sum \frac{\eta_j}{\eta_j + \lambda}}{\sigma^2_{\varepsilon}\sqrt{2\sum \left(\frac{\eta_j}{\eta_j + \lambda}\right)^2}} = \frac{\sigma^2}{\sigma^2_{\varepsilon}}\cdot \frac{\text{tr}(K_{\lambda}K_V)}{\sqrt{2\sum \left(\frac{\eta_j}{\eta_j + \lambda}\right)^2}}\\
\mathbb{E}T_{\lambda}^* &= \frac{\sigma^2}{\sigma^2_{\varepsilon}}\cdot \frac{\sum_{j = 1}^p \frac{\eta_j^2}{\eta_j + \lambda}}{\sqrt{2\sum \left(\frac{\eta_j}{\eta_j + \lambda}\right)^2}}
\end{align}

(Trace identity for $\text{tr}(K_{\lambda}K_V)$ shown below.)

\subsubsection{Asymptotic Equivalences}

For $n$ sufficiently large, we can assume the expected values of the test statistics are

\begin{align*}
\mathbb{E}T_F^* &= \frac{\mathbb{E}T_F - \sigma_{\varepsilon}^2p}{\sigma^2_{\varepsilon}\sqrt{2p}} = \frac{\sigma^2}{\sigma^2_{\varepsilon}}\cdot n\sqrt{\frac{p}{2}}\\
\mathbb{E}T_R^* &= \frac{\mathbb{E}T_R - \sigma_{\varepsilon}^2np}{\sigma^2_{\varepsilon}\sqrt{2\sum \eta_j^2}} = \frac{\sigma^2}{\sigma^2_{\varepsilon}}\sqrt{\frac{1}{2}\sum\eta_j^2}\\
\mathbb{E}T_{\lambda}^* &= \frac{\sigma^2}{\sigma^2_{\varepsilon}}\cdot \frac{\sum_{j = 1}^p \frac{\eta_j^2}{\eta_j + \lambda}}{\sqrt{2\sum \left(\frac{\eta_j}{\eta_j + \lambda}\right)^2}}
\end{align*}

Of course for $\lambda = 0$ we have $\mathbb{E}T_{\lambda}^* = \mathbb{E}T_F^*$ as usual.  Considering the limit as $\lambda \to \infty$ of the square of the second quotient in $T_{\lambda}^*$ (which is the only part that depends on $\lambda$):

\begin{align*}
\lim_{\lambda \to \infty} \frac{\left(\sum_{j = 1}^p \frac{\eta_j^2}{\eta_j + \lambda}\right)^2}{2\sum \left(\frac{\eta_j}{\eta_j + \lambda}\right)^2}
&= \lim \frac{\frac{1}{1/\lambda}}{\frac{1}{1/\lambda}}\frac{\left(\sum_{j = 1}^p \frac{\eta_j^2}{\eta_j + \lambda}\right)^2}{2\sum \left(\frac{\eta_j}{\eta_j + \lambda}\right)^2}\\
&= \lim \frac{\left(\sum_{j = 1}^p \frac{\eta_j^2}{\eta_j/\lambda + 1}\right)^2}{2\sum \left(\frac{\eta_j}{\eta_j/\lambda + 1}\right)^2}\\
&= \frac{\left(\sum_{j = 1}^p \eta_j^2\right)^2}{2\sum \eta_j^2} = \left(\mathbb{E}T_{R}^*\right)^2\\
\Rightarrow \lim_{\lambda \to \infty} \mathbb{E}T_{\lambda}^* &= \mathbb{E}T_R^*
\end{align*}

Thus the expected value of the ridge test statistic $T_{\lambda}^*$ converges to that of the random effects test statistics $T_R^*$, and so the two tests have approximately equal power for sufficiently large $n$ and $\lambda$.

\subsubsection{Covariance of Mantel Test Statistics with Ridge Kernel}

For two quadratic forms $Y^T A_1 Y$ and $Y^T A_2 Y$ with $Y \sim N(\mu, \Sigma)$, the covariance is given by formula

$$\text{Cov}(Y^T A_1 Y, Y^T A_2 Y) = 2 \text{tr}(A_1 \Sigma A_2 \Sigma) + 4 \mu^T A_1\Sigma A_2 \mu.$$
Let $Z = U^T Y \sim N(0, \sigma^2 I_p)$ as above, and $\eta_{\lambda} = \text{diag}(\eta_j/(\eta_j + \lambda))$.  Consider two ridge Mantel test statistics $T_{\lambda_1}$ and $T_{\lambda_2}$.

\begin{align*}
\text{Cov}(Z^T\Delta_{\lambda_1}Z, Z^T\Delta_{\lambda_2}Z) &= 2\text{tr}\left(\Delta_{\lambda_1}(\sigma^2K_V + \sigma^2_{\varepsilon}I)\Delta_{\lambda_2}(\sigma^2K_V + \sigma^2_{\varepsilon}I)\right)\\
&= 2\left[(\sigma^2)^2\text{tr}\left(\Delta_{\lambda_1}K_V\Delta_{\lambda_2}K_V\right)\right.\\ &~~~~~~~~ \left.+ 2\sigma^2\sigma^2_{\varepsilon}\text{tr}\left(\Delta_{\lambda_1}\Delta_{\lambda_2}K_V\right) + (\sigma^2_{\varepsilon})^2\text{tr}\left(\Delta_{\lambda_1}\Delta_{\lambda_2}\right)\right]\\
&= 2\sum\left[(\sigma^2)^2 \frac{\eta_j^4}{(\eta_j + \lambda_1)(\eta_j + \lambda_2)} + 2\sigma^2\sigma^2_{\varepsilon} \frac{\eta_j^3}{(\eta_j + \lambda_1)(\eta_j + \lambda_2)}\right. \\ &\left.~~~~~~~~ + (\sigma^2_{\varepsilon})^2\frac{\eta_j^2}{(\eta_j + \lambda_1)(\eta_j + \lambda_2)}\right]\\
&= 2\sum \frac{\eta_j^2(\sigma^2\eta_j + \sigma^2_{\varepsilon})^2}{(\eta_j + \lambda_1)(\eta_j + \lambda_2)}
\end{align*}

Using this to compute $\text{Cov}(T_{\lambda_1}^*, T_{\lambda_2}^*)$:

\begin{align*}
\text{Cov}(T_{\lambda_1}^*, T_{\lambda_2}^*) &= \left[(\sigma^2_{\varepsilon})^2\sqrt{2\sum \left(\frac{\eta_j}{\eta_j + \lambda_1}\right)^2}\sqrt{2\sum \left(\frac{\eta_j}{\eta_j + \lambda_2}\right)^2}\right]^{-1} \cdot 2 \sum \frac{\eta_j^2(\sigma^2\eta_j + \sigma^2_{\varepsilon})^2}{(\eta_j + \lambda_1)(\eta_j + \lambda_2)}\\
&= \left(\frac{\sigma^2}{\sigma^2_{\varepsilon}}\right)^2\left[\sqrt{\sum \left(\frac{\eta_j}{\eta_j + \lambda_1}\right)^2}\sqrt{\sum \left(\frac{\eta_j}{\eta_j + \lambda_2}\right)^2}\right]^{-1} \sum \frac{\eta_j^2(\eta_j + \frac{\sigma^2_{\varepsilon}}{\sigma^2})^2}{(\eta_j + \lambda_1)(\eta_j + \lambda_2)}\\
&= \left(\frac{\sigma^2}{\sigma_{\varepsilon}^2}\right)^2\frac{\text{tr}\left(K_{\lambda_1}K_{\lambda_2}\left(K_V + \frac{\sigma_{\varepsilon}^2}{\sigma^2}\right)^2\right)}{\sqrt{\text{tr}(K_{\lambda_1}^2)}\sqrt{\text{tr}(K_{\lambda_2}^2)}}
\label{eqn:cov_T_lambdas}
\end{align*}

\subsubsection{Asymptotic Distribution of AdaMant Statistic}

By \cite{nadarajah2008exact}, the distribution of the maximum of two standard normal random variables with correlation $\rho$ is 

$$f(x) = 2 \phi(x) \Phi\left(\frac{1 - \rho}{\sqrt{1 - \rho^2}}x\right).$$

\begin{figure}[ht]
\centering
\includegraphics[scale = 0.7]{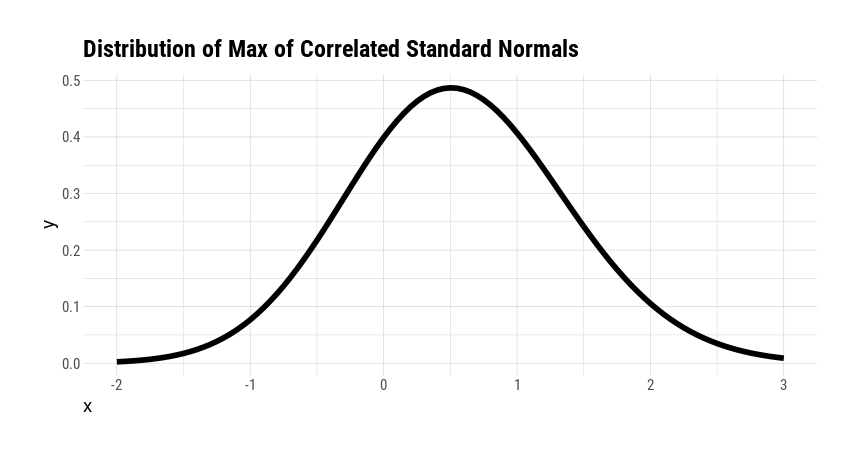}
\caption{Distribution of the maximum of two uncorrelated standard normal random variables.}
\label{fig:dist_max_uncorr_normals}
\end{figure}

Under the null distribution $H_0: \sigma^2 = 0$, giving $\rho = 0$, this reduces to

$$f_0(x) = 2\phi(x) \Phi(x),$$

displayed in Figure \ref{fig:dist_max_uncorr_normals}.

\subsection{Simulations}
\label{S-simulations}

\begin{figure}
\centering
\includegraphics[scale = 1]{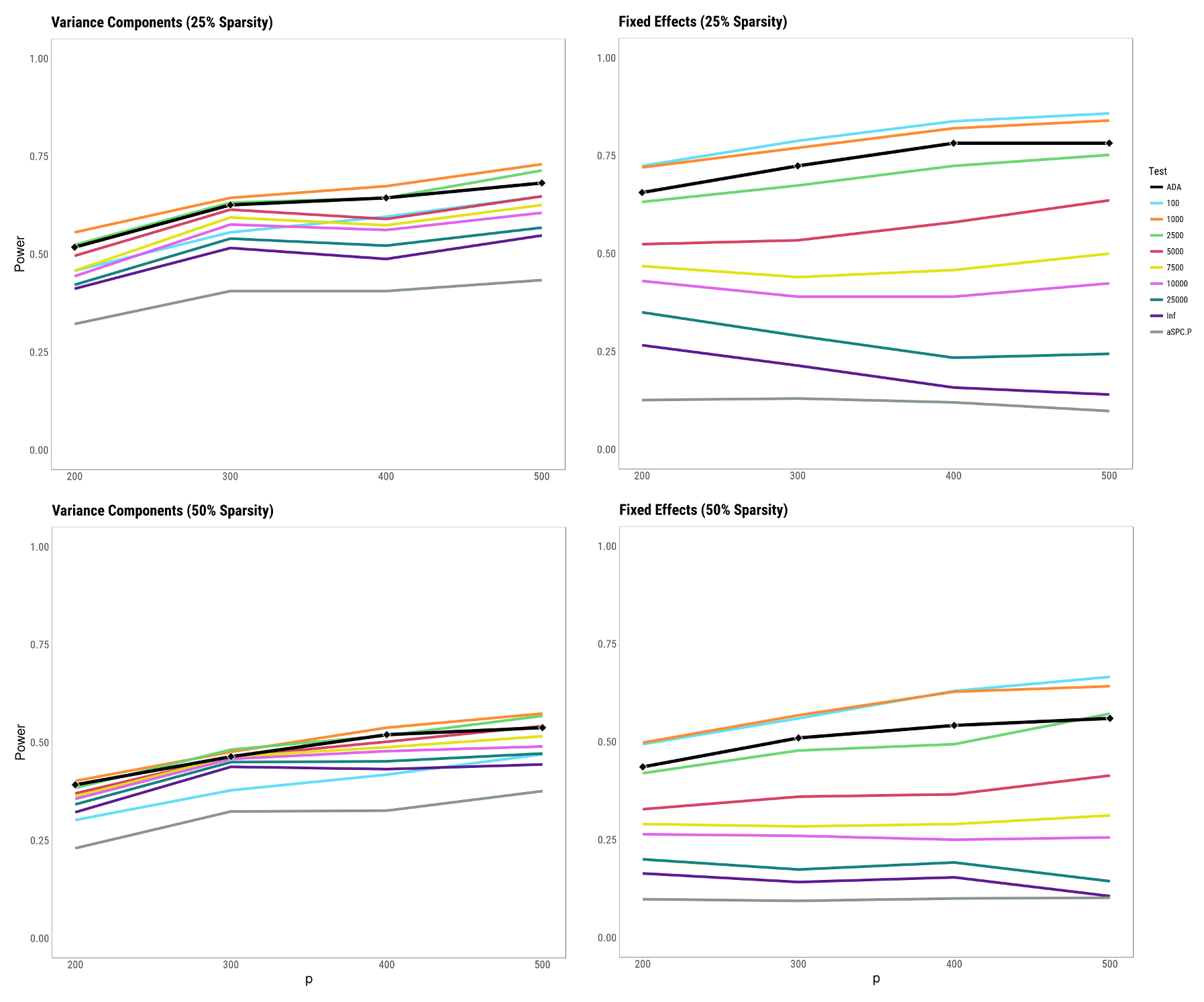}
\caption{Simulation results of the adaptive Mantel test for data generated from sparse models.  The sparsity level indicates the percentage of covariate effects that are set to 0.  The variance components models have $\sigma^2_b = 0.035$ for effective covariates; the fixed effects models have $\beta_j = (-1)^j 0.05$ for effective covariates.  All other simulation settings were the same those in Figure \ref{fig:sims}.  \textbf{(i)} Results for data generated from the variance components model with 25\% sparsity. \textbf{(ii)} Results for data generated from the variance components model with 50\% sparsity. \textbf{(iii)} Power for data generated from a fixed effects model with 25\% sparsity. \textbf{(iv)} Power for data generated from a fixed effects model with 50\% sparsity.}
\label{fig:sparsity_sim_plots}
\end{figure}

To assess the effect of unrelated covariates, we repeated the simulations from the variance components and fixed effects models with 25\% and 50\% of the covariate effects set to 0, (plots given in Figure \ref{fig:sparsity_sim_plots}). The shape of the plots and ordering of the curves with respect $\lambda$ are largely the same as in the 0\% sparsity case, with the power decreased according to level of sparsity as expected.  The power for AdaMant is reduced similar to the power for the single-parameter tests.  The variance components score test, dCov, and aSPC tests perform the worst of the tests considered, thus AdaMant is practical even when many included covariates are unrelated to the response variables.  In these cases, the results of AdaMant indicate that the use of the ridge kernel may exhibit greater power over these competing tests.

\newpage
\subsection{Applications}
\label{A-applications}

\subsubsection{Calculating Coherence}
\label{coherence}
\begin{enumerate}
  \item
  Suppose we have p channels and each channel has a time series $X(t)$ with length $T$.

  \item \textbf{Fourier transform}:
  We do discrete Fourier transform(DFT) on time series of each channel:
  \begin{center}
     $d(\omega_j)=T^{-1/2} \sum\limits^T_{t=1}X(t)\exp(-2\pi i\omega_j t)$
  \end{center}
  For $j=0,1,...T-1$, where the frequencies $\omega_j=j/T$. Denote the DFT of the k-th channel at frequency $\omega_j$ as $d(k,j)$. The cross spectrum of m,n-th channel is $d(m,j)\cdot d(n,j)^*$ (* denotes conjugate transpose) at frequency $\omega_j$, which is the m,n-th element of spectral matrix at frequency $\omega_j$.
  Denote the spectral matrix at frequency f as\\

\[
S(f) = \begin{pmatrix}
      S_{11}(f)  & S_{12}(f) &  \cdots & S_{1p}(f) \\
      S_{21}(f)  & S_{22}(f) &  \cdots & S_{2p}(f) \\
      \cdots    & \cdots &  \cdots & \cdots   \\
      S_{p1}(f)  & S_{p2}(f) &  \cdots & S_{pp}(f) \\
  \end{pmatrix}
\]

  \item \textbf{Averaging over frequencies within bands}: We set a cutoff value for frequency(0.146) which is about 45 Hz as the upper bound of gamma band. For each band, we average the spectral matrix over frequencies within that band.

  \item \textbf{Averaging over trials}: We average the spectrum matrix for each band over trials and get the 3-d array  ($channel\times channel\times bands$) which contains \textbf{spectral matrix} for each frequency band.
  \item \textbf{Calculate coherence matrix}: For each subject, We calculate the coherence matrix for each frequency band based on spectral matrices. The coherence between channel m and n can be calculated as
  \begin{equation*}
      r_{mn}(f)=\frac{|S_{mn}(f)|^2}{S_{mm}(f)S_{nn}(f)}
  \end{equation*}
\end{enumerate}

\subsubsection{EEG Experimental Design}

\begin{figure}[ht]
\centering
\includegraphics[scale = 0.25]{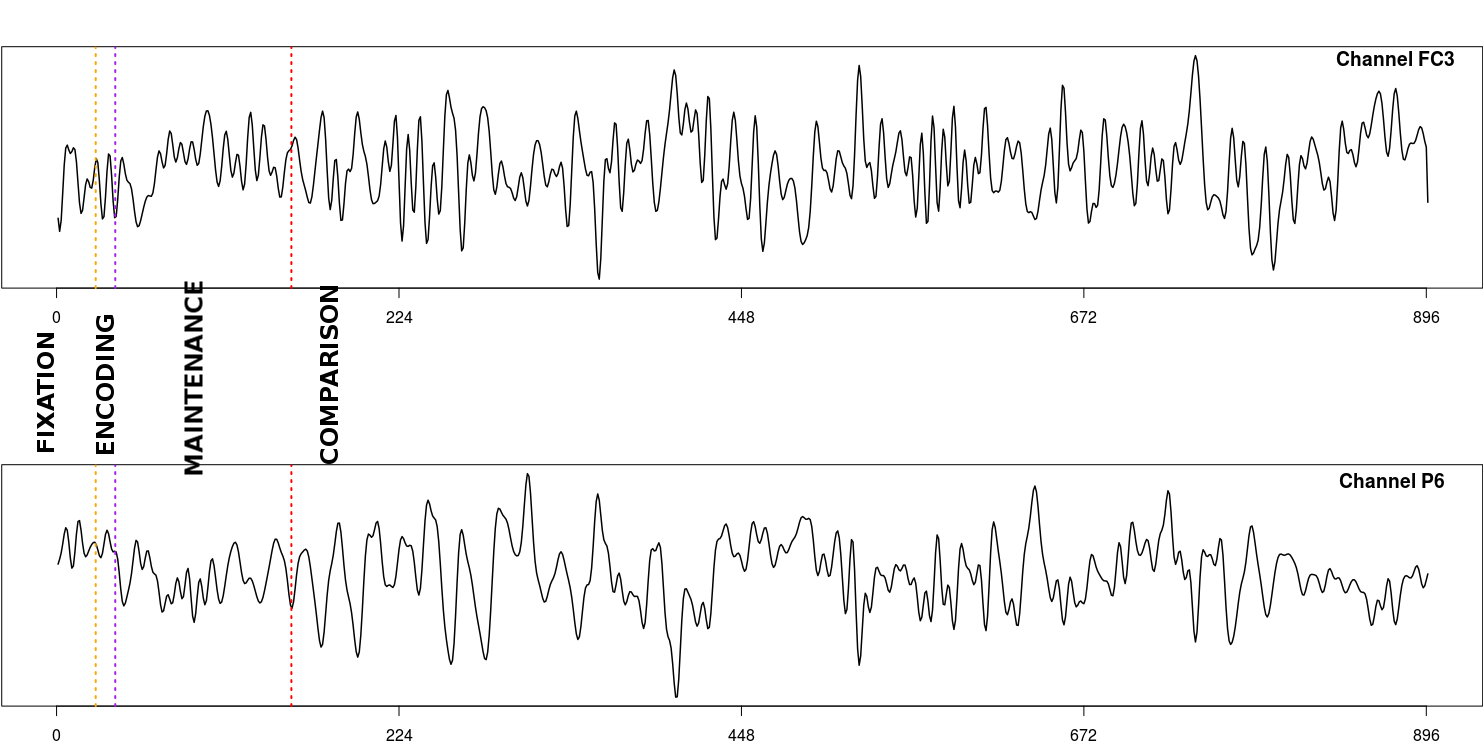}
\caption{Example of pre-processed EEG data from channels FC3 and P6 for one trial of the VWM experiment.}
\end{figure}

\begin{figure}[ht]
\centering
\includegraphics[scale = 0.95]{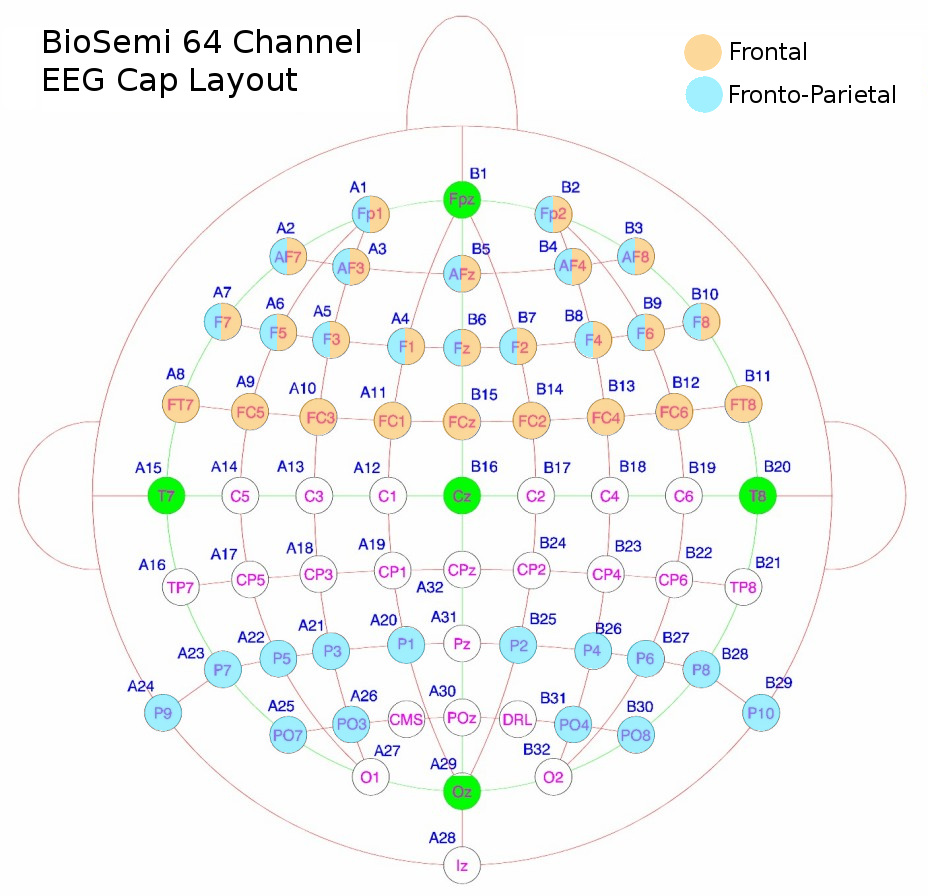}
\caption{Channel locations and labels for BioSemi 64 channel EEG cap.  Orange channels are those included in the "Frontal" channel group; blue channels are those included in the "Fronto-Parietal" channel group.}
\label{fig:biosemi_channel_groups}
\end{figure}

\subsubsection{EEG Analysis Additional Results}
\label{additional-results}

\begin{table}
\centering
\begin{tabular}{llrrr}
\hline
Band & Channels & $P$-value & $\lambda$ & $h^2$\\
\hline
Theta & Frontal & 0.0458 & $\infty$ & 0.165\\
Alpha & Frontal & 0.4100 & $\infty$ & 0.213\\
Beta & Frontal & 0.6890 & 0.1 & 0.191\\
Gamma & Frontal & 0.8860 & 10 & 0.187\\
Theta & All & 0.2610 & 0.1 & 0.178\\
Alpha & All	 & 0.8610 & 1 & 0.173\\
Beta & All & 0.3620 & $\infty$ & 0.221\\
Gamma & All & 0.4200 & $\infty$ & 0.184\\
\hline
\end{tabular}
\caption{Adaptive Mantel $P$-values for testing association of EEG coherence with genome wide similarity. Frontal channels include the 25 most anterior EEG channels.  The $\lambda$ column indicates the best ridge regression penalty chosen by AdaMant, with $\lambda = \infty$ corresponding to the variance components score test.}
\label{tab:gwas_mantel}
\end{table}

\begin{table}
\centering
\begin{tabular}{llrr}
\hline
Band & Channels & $P$-value & $\lambda$\\
\hline
Theta & Frontal & 0.085 & 0.5\\
Alpha & Frontal & 0.381 & 1\\
Theta & All & 0.416 & 0.5\\
Alpha & All & 0.065 & 5\\
\hline
\end{tabular}
\caption{AdaMant $P$-values for association of EEG Coherence for frontal and all channels with 11 selected AD-related SNPs.}
\end{table}

\begin{figure}[ht]
\centering
\includegraphics[scale = 0.5]{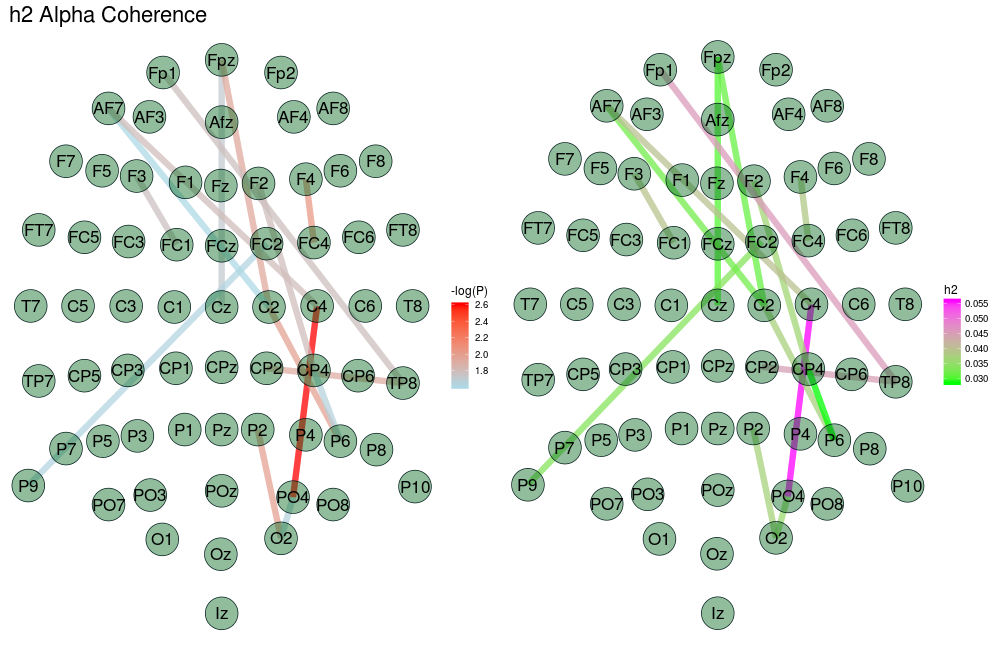}
\caption{Selection of channel pairs with alpha band coherence most strongly associated with AD SNPs.}
\label{fig:alpha_AD}
\end{figure}

To assess possible contributions of individual channel pairs, each channel pair was separately tested for significance with the AD SNPs with the adaptive Mantel test.  Figure \ref{fig:alpha_AD} shows the most significant channel pairs across all channel pairs for the alpha band.  The TP8 -- CP2 connection is the most significant at $P < 0.0001$.  Other top pairs are C4 -- PO4, C4 -- AF7, and T8 -- FT7.  The most significant connections are mostly between the right temporal regions with the left frontal regions.  For the theta band, the most significant channel pairs were P9 -- Cz, P9 -- CP3, CP3 -- AF3, and P8 -- F3.  The P9 channel also had relatively significant connections with many other channels in the frontal left hemisphere.  These results are supported by a number of previous studies that have established links between working memory performance and features measured by EEG.  For instance, \citeauthor{onton2005frontal} (\citeyear{onton2005frontal}) found increases in frontal midline theta power with increasing memory load during a verbal-working memory task;  \citeauthor{sauseng2005fronto} (\citeyear{sauseng2005fronto}) also found that alpha coherence plays a significant role in ``top-down'' control during working memory tasks; and \citeauthor{simons2003prefrontal} (\citeyear{simons2003prefrontal}) identified important interactions between the prefrontal and medial temporal lobes for the processing of long-term memory.

Similarly, the association of each individual SNP with EEG coherence was assessed with AMT.  For alpha coherence with all channels, the most significant SNP ($P = 0.02$) is rs2227564, a functional polymorphism within plasminogen activator urokinase (PLAU) gene.  An allele of this SNP has been linked to significantly higher plaque counts in AD, although its role is not well-established.  The second most significant SNP from the individual tests is rs3851179, a SNP upstream of the PICALM gene.  This SNP has been repeatedly implicated as a factor in AD \parencite{harold2009genome}, as well as Parkinson's Disease and schizophrenia \parencite{zhou2010association}, although there are dissenting results regarding its significance in particular Chinese populations \parencite{chen2012polymorphisms, liu2013lack}.  In the genetics literature on cognitive function in healthy subjects, polymorphisms in neurotransmitter genes, such as those in the dopamine pathway, have been shown to be significantly associated with increased neuronal activity in the prefrontal cortex during working memory tasks, as measured by fMRI \parencite{Bertolino3918}.  \citeauthor{vogler2014substantial} (\citeyear{vogler2014substantial}) analyzed data from the $n$-back memory task for 2298 subjects, and estimated genome-wide heritability of working memory accuracy to be 41\% (95\% CI: 0.13, 0.69).  Taken together, these results make it plausible that genetic factors have an important influence on brain function related to working memory, but determining the specific nature of the role of genetics in brain function remains a challenging problem that will require repeated validation through a variety of different studies and experiments. 

\end{document}